\documentclass[12pt,twocolumn]{IEEEtran}
\usepackage[T1]{fontenc}
\usepackage[latin9]{inputenc}
\usepackage{color}
\usepackage{array}
\usepackage{float}
\usepackage{booktabs}
\usepackage{textcomp}
\usepackage{amsmath}
\usepackage{amsthm}
\usepackage{amssymb}
\usepackage{stackrel}
\usepackage{graphicx}
\usepackage{setspace}
\usepackage{wasysym}
\usepackage[unicode=true,
 bookmarks=true,bookmarksnumbered=true,bookmarksopen=true,bookmarksopenlevel=1,
 breaklinks=false,pdfborder={0 0 0},pdfborderstyle={},backref=false,colorlinks=false]
 {hyperref}
\hypersetup{pdftitle={Your Title},
 pdfauthor={Your Name},
 pdfpagelayout=OneColumn, pdfnewwindow=true, pdfstartview=XYZ, plainpages=false}

\makeatletter

\providecommand{\tabularnewline}{\\}
\floatstyle{ruled}
\newfloat{algorithm}{tbp}{loa}
\providecommand{\algorithmname}{Algorithm}
\floatname{algorithm}{\protect\algorithmname}

\let\oldforeign@language\foreign@language
\DeclareRobustCommand{\foreign@language}[1]{%
  \lowercase{\oldforeign@language{#1}}}
\theoremstyle{plain}
\newtheorem{thm}{\protect\theoremname}
\theoremstyle{plain}
\newtheorem{lem}[thm]{\protect\lemmaname}
\theoremstyle{plain}
\newtheorem{cor}[thm]{\protect\corollaryname}

\usepackage[caption=false,font=footnotesize]{subfig}

\makeatother

\providecommand{\corollaryname}{Corollary}
\providecommand{\lemmaname}{Lemma}
\providecommand{\theoremname}{Theorem}

\begin{document}
\title{Large-Scale Rate-Splitting Multiple Access in Uplink UAV Networks:
Effective Secrecy Throughput Maximization Under Limited Feedback Channel}
\author{Hamed~Bastami,~Hamid~Behroozi,~\IEEEmembership{Member,~IEEE,}~Majid~Moradikia,~~Ahmed~Abdelhadi,~\\
Derrick~Wing~Kwan~Ngand,~\IEEEmembership{Fellow,~IEEE},~and
Lajos~Hanzo,~\IEEEmembership{Life~Fellow,~IEEE}\thanks{Hamed~Bastami and Hamid~Behroozi are with the Department of Electrical
Engineering, Sharif University of Technology, Tehran, Iran, e-mail{\small{}s:
\{hamed.bastami@ee., behroozi@\}sharif.edu}}\thanks{Majid Moradikia is with Department of Data Science Worcester Polytechnic
Institute, Worcester, Massachusetts, e-mail: \{mmoradikia@wpi.edu\}.
Ahmed Abdelhadi is with the Engineering Technology Department at University
of Houston, e-mail{\small{}s:\{aabdelhadi\}@uh.edu}.}\thanks{Derrick Wing Kwan N is with the School of Electrical Engineering and
Telecommunications, University of New South Wales, Sydney, Australia,
e-mail: \protect\href{mailto:w.k.ng@unsw.edu.au}{w.k.ng@unsw.edu.au}.}\thanks{Lajos Hanzo is with the University of Southampton, Southampton SO17
1BJ, U.K, e-mail: \protect\href{mailto:hanzo@soton.ac.uk}{hanzo@soton.ac.uk}.}}

\maketitle
\begin{abstract}
Unmanned aerial vehicles (UAVs) are capable of improving the performance
of next generation wireless systems. Specifically, UAVs can be exploited
as aerial base-stations (UAV-BS) for supporting legitimate ground
users in remote uncovered areas or in environments temporarily requiring
high capacity. However, their communication performance is prone to
both channel estimation errors and potential eavesdropping. Hence,
we investigate the effective secrecy throughput of the UAV-aided uplink,
in which rate-splitting multiple access (RSMA) is employed by each
legitimate user for secure transmission under the scenario of massive
access. To maximize the effective network secrecy throughput in the
uplink, the transmission rate vs. power allocation relationship is
formulated as a max-min optimization problem, relying on realistic
imperfect channel state information (CSI) of both the legitimate users
and of the potential eavesdroppers ($Eves$). We then propose a novel
transformation of the associated probabilistic constraints for decoupling
the variables, so that our design problem can be solved by alternatively
activating the related block coordinate decent programming. In the
model considered, each user transmits a superposition of two messages
to a UAV-BS, each having different transmit power and the UAV-BS uses
a successive interference cancellation (SIC) technique to decode the
received messages. Given the non-convexity of the problem, it is decoupled
into a pair of sub-problems. In particular, we derive a closed form
expression for the optimal rate-splitting fraction of each user. Then,
given the optimal rate-splitting fraction of each user, the $\epsilon$-constrainted
transmit power of each user is calculated by harnessing sequential
parametric convex approximation (SPCA) programming. Finally, the optimal
SIC order is determined by an exhaustive search method. Our simulation
results confirm that the scheme conceived significantly improves the
effective secrecy throughput compared to both the existing orthogonal
and non-orthogonal benchmarks as well as to the RSMA scheme ignoring
CSI uncertainty.
\end{abstract}


\IEEEpeerreviewmaketitle{}

\section{Introduction}

\IEEEPARstart{I}{n} order to support the emerging Beyond 5G (B5G)
system concept, unmanned aerial vehicles (UAV) may be harnessed as
air-borne base-station (BS), particularly in areas of high tele-traffic
density \cite{=00005B1=00005D}-\cite{=00005B5=00005D}. However,
owing to their LoS propagation UAV-BSs usually suffer from strong
co-channel interference. Although this problem can be potentially
mitigated by the sophisticated trajectory design of UAVs \cite{=00005B1=00005D,=00005B1-1=00005D},
the degree of freedom attained is typically inadequate to support
the ever-growing terrestrial user population. In this context, multiple
access (MA) techniques play a crucial role in fulfilling the high
data rate, low latency, and massive connectivity requirements, as
the three most important Key Performance Indicators (KPI)s for B5G
\cite{=00005B2=00005D}-\cite{=00005B12=00005D}. 

Rate-splitting multiple access (RSMA) has attracted a great deal of
interest, as a key-enabling radio access technology capable of satisfying
the massive connectivity requirements of B5G \cite{=00005B2=00005D,=00005B4=00005D,=00005B5=00005D},
\cite{=00005B8=00005D}-\cite{=00005B12=00005D}. Briefly, RSMA is
a generalization of non-orthogonal multiple access (NOMA) and space-division
multiple access (SDMA) \cite{=00005B8=00005D}, that outperforms both
schemes in terms of its robustness and spectral efficiency. In a rate-splitting
(RS) scheme, the transmitted signals are split into two parts at the
transmitter (Tx), namely into a common message and a private message.
Subsequently, by performing successive interference cancellation (SIC)
at the receiver (Rx), the capacity region of the MA channel (MAC)
can be approached. Inspired by this promising MA framework, most of
the RSMA-based literature considered the downlink (DL) \cite{=00005B2=00005D,=00005B4=00005D,=00005B5=00005D,=00005B8=00005D,=00005B9=00005D}
even though the DL actually represents a broadcast scenario and multiple
access is only possible in the uplink (UL). In \cite{=00005B10=00005D}
a RS scheme was designed for guaranteeing max-min fairness in UL-NOMA.
As a further advance, a cooperative rate-splitting (CRS) UL scheme
was proposed in \cite{=00005B11=00005D}, where each user broadcasts
his/her signal during the first phase and receives the transmitted
signal of the other user, while during the second phase, each user
relays the other user\textquoteright s message. Then, Yang \textit{et
al}. \cite{=00005B12=00005D}, proposed UL-RSMA for maximizing the
users\textquoteright{} sum-rate sum-rate by optimally sharing the
total transmit powers of both user-messages, while exhaustively searching
for the optimal decoding order at the SIC receiver. However, these
contributions stipulate the idealized simplifying assumption of having
perfect channel state information (CSI) for resource allocation design,
which is not realistic in practice. More particularly, in massive
access scenarios in which a large number of CSIs have to be reported
to the BS using limited feedback having CSI imperfections is unavoidable,
resulting in link outage \cite{=00005B7=00005D,=00005B13=00005D}.
More importantly, none of the above-mentioned RSMA UL scenarios of
\cite{=00005B10=00005D}-\cite{=00005B12=00005D} have addressed the
associated security concerns. In particular, the concurrent UL transmissions
of a massive number of messages over the same bandwidth increases
the risk of security breaches. To protect the confidentiality of the
transmitted signals, physical layer security (PLS) techniques can
be exploited for increasing the channel capacity difference between
the legitimate and eavesdropping links. Unfortunately, MA systems
are particularly susceptible to passive attacks, since the eavesdroppers
($Eves$) have more target users, they can glean information from
\cite{=00005B4=00005D,=00005B7=00005D,=00005B9=00005D}. In this context,
jamming aims for confusing the potential $Eves$ by deliberately injecting
specifically designed artificial noise (AN) with the aid of beamforming
\cite{=00005B14=00005D,=00005B15=00005D}. As a further development,
the authors of \cite{=00005B16=00005D,=00005B17=00005D} proposed
a secure NOMA approach. However, since the super-imposed non-orthogonal
signals may be detected by SIC at $Eve$, superposition potentially
degrades the level of security. In fact, after detecting the superimposed
streams $Eve$ becomes capable of wiretapping the rest of the embedded
information, which is becoming less interference-infested. By contrast,
using an RSMA scheme, the common message plays the dual roles of the
desired message as well as that of the AN without the need for assigning
a portion of the limited transmit power to the AN \cite{=00005B4=00005D,=00005B5=00005D,=00005B9=00005D}.
However, the secure RSMA designs of \cite{=00005B4=00005D,=00005B5=00005D,=00005B9=00005D}
considered the DL scenario, hence their results are not applicable
to the UL due to the different nature of the problems. To the best
of our knowledge, at the time of writing,  no attention has been devoted
to the integration of UL-RS with UAV-BS. Furthermore, the robust and
secure design of RSMA-aided UAV networks relying on realistic imperfect
CSI has not been investigated so far. 

Given the knowledge gaps mentioned above, we consider a network in
which the legitimate users aim for communicating with a UAV-BS in
the presence of multiple passive $Eve$. In this UL scenario, each
user employs RS, where the corresponding message of each user is split
into two parts. Then, each user transmits a superposition of two messages
having different power levels. To realize massive connectivity, we
assume furthermore that at the network initialization a clustering
process is accomplished by which the users are divided into different
non-overlapping groups. Furthermore, due to the limited CSI feedback
accuracy, a link outage may occur. Hence we introduce a maximum tolerable
connection outage probability (COP) constraint for quantifying its
impact on the system performance. It is worth mentioning that in contrast
to the RSMA downlink in \cite{=00005B4=00005D,=00005B9=00005D}, where
the authors considered the secure design of the common streams, here
we exploit a different strategy, where the transmission rate and the
power allocated to each part of the bipartite messages is optimized
in terms of \textit{Effective Network Secrecy Throughput} (ENST) maximization.
ENST is a secrecy performance metric quantifying the average secure
throughput. More explicitly, when the reception reliability is considered
to be similarly important to the security, then this parameter is
considered. Mathematically, ENST is formulated as the product of the
target secrecy rate and the probability of successful reception as
defined in \cite[Eq. (5)]{=00005B31=00005D}. Against this background,
our contributions are summarized as follows: 
\begin{itemize}
\item In addition to the COP constraint, which captures the impact of link
outages, the secrecy outage probability (SOP) is tightly controlled
to be under the tolerable level under unknown CSI of the $Eve$. We
then maximize the ENST, subject to both COP and SOP constraints, as
well as to the limited power budget. In particular, we design the
RS power allocation at the users as well as the SIC-ordering corresponding
to each cluster, so that the ENST is maximized. 
\item To deal with the resultant non-convex problem, we first derive a closed-form
expression for characterizing the COP and a tight approximation of
the SOP constraints. Then, we harness the two-tier block coordinate
decent technique, where the optimization variables are estimated successively
in an iterative manner. The first loop of this twin-tier approach
maximizes the transmission rates, leading to a closed-form optimal
solution relying on the Lambert $W$-function. By contrast, the second
loop encounters some non-convexities, which are tackled by the powerful
sequential parametric convex approximation (SPCA) method. The convex
approximation of the non-convex factors are found with the aid of
the first-order Taylor expansion.
\item Our simulation results demonstrate that the proposed framework outperforms
the existing non-orthogonal benchmarks in terms of the ENST criterion.
\begin{table*}[tbh]
\caption{\textcolor{black}{\footnotesize{}Boldly and explicitly contrasting
our contributions to the existing literature.}}
\begin{tabular}{|>{\raggedright}p{4cm}|>{\centering}m{1.3cm}|>{\centering}m{0.2cm}|>{\centering}m{0.2cm}|>{\centering}m{0.2cm}|>{\raggedright}m{0.2cm}|>{\centering}m{0.2cm}|>{\raggedright}m{0.2cm}|>{\raggedright}m{0.2cm}|>{\raggedright}m{0.2cm}|>{\raggedright}m{0.2cm}|>{\raggedright}m{0.2cm}|>{\raggedright}m{0.2cm}|>{\raggedright}m{0.2cm}|>{\raggedright}m{0.2cm}|>{\raggedright}m{0.2cm}|>{\raggedright}m{0.2cm}|>{\raggedright}m{0.2cm}|>{\raggedright}m{0.2cm}|>{\raggedright}m{0.3cm}|}
\hline 
{\small{}}%
\begin{tabular}{c}
{\small{}References$\Rightarrow$}\tabularnewline
\hline 
\hline 
{\small{}Keywords$\Downarrow$}\tabularnewline
\end{tabular} & \textbf{\footnotesize{}Our Approach} & {\footnotesize{}\cite{=00005B1=00005D}} & {\footnotesize{}\cite{=00005B1-1=00005D}} & {\footnotesize{}\cite{=00005B2=00005D}} & {\footnotesize{}\cite{=00005B3=00005D}} & {\footnotesize{}\cite{=00005B4=00005D}} & {\footnotesize{}\cite{=00005B5=00005D}} & {\footnotesize{}\cite{=00005B6=00005D}} & {\footnotesize{}\cite{=00005B7=00005D}} & {\footnotesize{}\cite{=00005B8=00005D}} & {\footnotesize{}\cite{=00005B9=00005D}} & {\footnotesize{}\cite{=00005B10=00005D}} & {\footnotesize{}\cite{=00005B11=00005D}} & {\footnotesize{}\cite{=00005B12=00005D}} & {\footnotesize{}\cite{=00005B13=00005D}} & {\footnotesize{}\cite{=00005B14=00005D}} & {\footnotesize{}\cite{=00005B15=00005D}} & {\footnotesize{}\cite{=00005B16=00005D}} & {\footnotesize{}\cite{=00005B17=00005D}}\tabularnewline
\hline 
\hline 
\textbf{\footnotesize{}UAV-BS} & \textbf{$\checkmark$} & \textbf{$\checkmark$} & \textbf{$\checkmark$} & \textbf{$\checkmark$} & \textbf{$\checkmark$} & $\checkmark$ & \textbf{$\checkmark$} &  &  &  &  &  &  &  &  &  &  &  & \tabularnewline
\hline 
\textbf{\footnotesize{}UAV Trajectory Design} &  & \textbf{$\checkmark$} & \textbf{$\checkmark$} &  &  &  &  &  &  &  &  &  &  &  &  &  &  &  & \tabularnewline
\hline 
\textbf{\footnotesize{}IUI Management} &  & \textbf{$\checkmark$} & \textbf{$\checkmark$} &  &  & \textbf{$\checkmark$} & \textbf{$\checkmark$} & \textbf{$\checkmark$} &  &  &  &  &  &  &  &  &  &  & \tabularnewline
\hline 
\textbf{\footnotesize{}IUI Cancellation} & \textbf{$\checkmark$} &  &  &  &  &  &  &  &  &  &  &  &  & \textbf{$\checkmark$} &  &  &  &  & \tabularnewline
\hline 
\textbf{\footnotesize{}SIC Ordering} & \textbf{$\checkmark$} &  &  &  &  &  &  &  &  &  &  &  &  & \textbf{$\checkmark$} &  &  &  &  & \tabularnewline
\hline 
\textbf{\footnotesize{}RSMA} & \textbf{$\checkmark$} &  &  & \textbf{$\checkmark$} &  & \textbf{$\checkmark$} & \textbf{$\checkmark$} &  &  & \textbf{$\checkmark$} & \textbf{$\checkmark$} & \textbf{$\checkmark$} & \textbf{$\checkmark$} & \textbf{$\checkmark$} &  &  &  &  & \tabularnewline
\hline 
\textbf{\footnotesize{}NOMA} &  &  &  &  &  &  &  &  & \textbf{$\checkmark$} &  &  &  &  &  &  &  &  & \textbf{$\checkmark$} & \textbf{$\checkmark$}\tabularnewline
\hline 
\textbf{\footnotesize{}SDMA} &  &  &  &  &  &  &  & \textbf{$\checkmark$} &  &  &  &  &  &  &  &  &  &  & \tabularnewline
\hline 
\textbf{\footnotesize{}DL} &  &  &  & \textbf{$\checkmark$} &  & \textbf{$\checkmark$} & \textbf{$\checkmark$} &  & \textbf{$\checkmark$} & \textbf{$\checkmark$} & \textbf{$\checkmark$} &  &  &  &  &  &  & \textbf{$\checkmark$} & \textbf{$\checkmark$}\tabularnewline
\hline 
\textbf{\footnotesize{}UL} & \textbf{$\checkmark$} &  &  &  &  &  &  &  &  &  &  & \textbf{$\checkmark$} & \textbf{$\checkmark$} & \textbf{$\checkmark$} &  &  &  & \textbf{$\checkmark$} & \tabularnewline
\hline 
\textbf{\footnotesize{}Limited Feedback Error} & \textbf{$\checkmark$} &  &  &  &  &  &  & \textbf{$\checkmark$} & \textbf{$\checkmark$} &  &  &  &  &  & \textbf{$\checkmark$} &  &  &  & \tabularnewline
\hline 
\textbf{\footnotesize{}Imperfect CSI} & \textbf{$\checkmark$} &  &  &  &  & \textbf{$\checkmark$} & \textbf{$\checkmark$} &  & \textbf{$\checkmark$} &  & \textbf{$\checkmark$} &  &  &  & \textbf{$\checkmark$} &  &  & \textbf{$\checkmark$} & \tabularnewline
\hline 
\textbf{\footnotesize{}PLS} & \textbf{$\checkmark$} &  &  &  &  & \textbf{$\checkmark$} & \textbf{$\checkmark$} &  & \textbf{$\checkmark$} &  & \textbf{$\checkmark$} &  &  &  & \textbf{$\checkmark$} & \textbf{$\checkmark$} & \textbf{$\checkmark$} & \textbf{$\checkmark$} & \textbf{$\checkmark$}\tabularnewline
\hline 
\textbf{\footnotesize{}AN Design} &  &  &  &  &  & \textbf{$\checkmark$} &  &  &  &  &  &  &  &  &  & \textbf{$\checkmark$} & \textbf{$\checkmark$} &  & \tabularnewline
\hline 
\textbf{\footnotesize{}Beamformer (precoder) Design} & \textbf{$\checkmark$} &  &  &  &  & \textbf{$\checkmark$} & \textbf{$\checkmark$} &  &  &  & \textbf{$\checkmark$} &  &  &  &  & \textbf{$\checkmark$} & \textbf{$\checkmark$} & \textbf{$\checkmark$} & \textbf{$\checkmark$}\tabularnewline
\hline 
\textbf{\footnotesize{}Power Allocation} & \textbf{$\checkmark$} &  &  &  &  & \textbf{$\checkmark$} & \textbf{$\checkmark$} &  & \textbf{$\checkmark$} &  & \textbf{$\checkmark$} &  &  &  &  & \textbf{$\checkmark$} & \textbf{$\checkmark$} & \textbf{$\checkmark$} & \tabularnewline
\hline 
\textbf{\footnotesize{}Known $Eve$ with Imperfect E-CSIT} &  &  &  &  &  & \textbf{$\checkmark$} & \textbf{$\checkmark$} &  &  &  & \textbf{$\checkmark$} &  &  &  &  &  &  & \textbf{$\checkmark$} & \tabularnewline
\hline 
\textbf{\footnotesize{}Worst-Case Secrecy Rate Maximization} &  &  &  &  &  & \textbf{$\checkmark$} & \textbf{$\checkmark$} &  &  &  & \textbf{$\checkmark$} &  &  &  &  &  &  &  & \tabularnewline
\hline 
\textbf{\footnotesize{}Max-Min Fairness} &  &  &  &  &  & \textbf{$\checkmark$} & \textbf{$\checkmark$} &  &  &  & \textbf{$\checkmark$} & \textbf{$\checkmark$} &  &  &  &  &  &  & \tabularnewline
\hline 
\textbf{\footnotesize{}Sum-Rate Maximization} &  &  &  &  &  &  &  &  & \textbf{$\checkmark$} &  &  &  &  & \textbf{$\checkmark$} &  &  &  &  & \tabularnewline
\hline 
\textbf{\footnotesize{}Known $Eve$} &  &  &  &  &  &  &  &  &  &  &  &  &  &  &  &  & \textbf{$\checkmark$} &  & \tabularnewline
\hline 
\textbf{\footnotesize{}Unknown $Eve$} & \textbf{$\checkmark$} &  &  &  &  &  &  &  &  &  &  &  &  &  &  & \textbf{$\checkmark$} &  &  & \tabularnewline
\hline 
\textbf{\footnotesize{}ICI Cancellation} & \textbf{$\checkmark$} &  &  &  &  &  &  &  &  &  &  &  &  &  &  &  &  &  & \tabularnewline
\hline 
\textbf{\footnotesize{}ENST Maximization} & \textbf{$\checkmark$} &  &  &  &  &  &  &  &  &  &  &  &  &  &  &  &  &  & \tabularnewline
\hline 
\textbf{\footnotesize{}COP Constraint} & \textbf{$\checkmark$} &  &  &  &  &  &  &  &  &  &  &  &  &  &  &  &  &  & \tabularnewline
\hline 
\textbf{\footnotesize{}SOP Constraint} & \textbf{$\checkmark$} &  &  &  &  &  &  &  &  &  &  &  &  &  &  &  &  &  & \tabularnewline
\hline 
\end{tabular}
\end{table*}
\end{itemize}
Our contributions are boldly and explicitly contrasted to the state-of-the-art
at a glance in Table $1$. The rest of this paper is organized as
follows. The system model and channel definitions are provided in
Section II. Section III describes the signal representation and formulates
our ENST maximization problem. The proposed SPCA-based solution, the
two-tier block coordinate decent procedures and our complexity analysis
are provided in Section IV. In Section V, our simulation results are
presented and the paper is concluded in Section VI. Finally, the Appendices
and Proofs of the claims are provided in Section VII. 

\textit{Notation:} Vectors and matrices are denoted by lower-case
and upper-case boldface symbols, respectively; $\left(.\right)^{\mathrm{T}}$,
$\left(.\right)^{\mathrm{*}}$, $\left(.\right)^{\mathrm{H}}$, and
$\left(.\right)^{\mathrm{-1}}$ denote the transpose, conjugate, conjugate
transpose, and inverse of a matrix respectively; $\mathfrak{R}e(.)$
denotes the real part of a complex variable, and $\mathfrak{I}m(.)$
the imaginary part of a complex variable; We use $\mathbb{E}{\left\{ \cdot\right\} }$
and $\triangleq$ to denote the expectation operation and a definition,
respectively. A complex Gaussian random variable with mean $\mathit{\mu}$
and variance $\sigma^{2}$ reads as $\mathcal{C}\mathcal{N}\left(\mathit{\mu},\sigma^{2}\right)$,
and $\textrm{Exp}\left(\lambda\right)$, $\textrm{Beta}\left(\alpha,\beta\right)$,
and $\textrm{Gamma}\left(\gamma,\zeta\right)$ respectively denote
the exponential distribution with mean $\lambda$, beta-distribution
with parameters $\alpha$ and $\beta$, and gamma-distribution with
shape $\gamma$ and rate $\zeta$. The principal branch of the Lambert
$W$-function is defined by $W_{0}\left(x\right)e^{W_{0}\left(x\right)}=x$
for $x\ge-\frac{1}{e}$ with $W_{0}\left(x\right)\ge-1$ \cite{=00005B18=00005D};
$\mathbf{I}_{N}$ denotes the $N\times N$ identity matrix; $\mathbb{R}^{N\text{\texttimes}1}$
and $\mathbb{C}^{N\times1}$ denote the set of $N$-dimensional standard
real and complex Gaussian random variable, respectively; $\mathbb{C}^{N\times N}$
stands for an $N\times N$ element standard complex Gaussian random
matrix whose real and imaginary parts are independent normally distributed
random variables with a mean of zero and variance $\frac{1}{2}$.
The notations $\left[x\right]^{+}$ and $\mathbb{P}\left(.\right)$
stand for $\textrm{max}{\left\{ x,0\right\} }$ and probability, respectively.
The entry in the $i$-th row of a vector $\mathbf{h}$ is represented
by $\mathbf{h}\left[i\right]$. Furthermore, $\mathbf{u}^{max}\left\{ \mathbf{A}\right\} $
and $\mathbf{v}^{max}\left\{ \mathbf{A}\right\} $ denote the columns
of $\mathbf{U}_{\mathbf{A}}$ and $\mathbf{V}_{\mathbf{A}}$ corresponding
to the dominant singular value $\lambda^{max}\left\{ \mathbf{A}\right\} $
of matrix $\mathbf{A}$, respectively, i.e., the matrix $\mathbf{A}$
has a Singular Value Decomposition (SVD) given by $\mathbf{A}\triangleq\mathbf{U}_{\mathbf{A}}\mathbf{\mathbf{\Lambda}}_{\mathbf{A}}\mathbf{V}_{\mathbf{A}}$.
Finally, $\angle\left(\mathbf{u},\mathbf{v}\right)$ represents the
angle between vectors $\mathbf{v}$ and $\mathbf{u}$.

\section{System Model}

\begin{figure}[tbh]
\begin{onehalfspace}
\centering{}\includegraphics[viewport=5bp 5bp 1401bp 835bp,clip,scale=0.18]{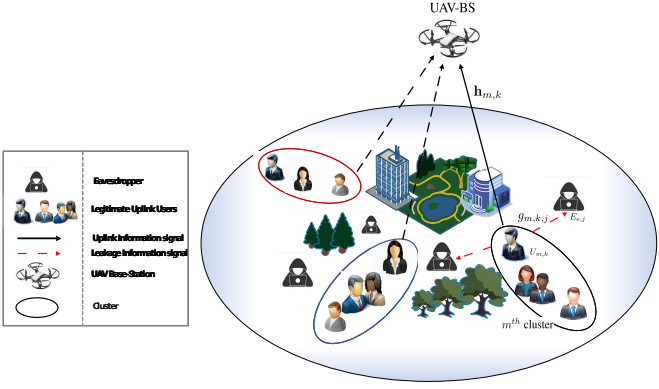}\caption{\label{fig:Proposed-System-Model}The considered system model adopting
RSMA}
\end{onehalfspace}
\end{figure}

We consider the secure single-input multi-output (SIMO) uplink system,
shown in Fig. \ref{fig:Proposed-System-Model}. There are $M$ clusters
in the network considered, whose $m^{th}$ cluster includes $K_{m}$
number of single-antenna legitimate users gathered in the set $\boldsymbol{U}_{m}\triangleq\left\{ U_{m,k}\right\} $,
$\forall k\in\mathcal{K}_{m}\triangleq\left\{ 1,\ldots,K_{m}\right\} $,
who aim for transmitting to an $N_{t}$-antenna UAV-BS. We consider
a massive access setting, where $\sum_{m=1}^{M}K_{m}\gg M$. Meanwhile,
$J$ number of non-cooperative passive $N_{e}$-antenna eavesdroppers
($Eves$) gathered in the set $\boldsymbol{E}\triangleq\left\{ E_{e,j}\right\} ,\forall j\in\mathcal{J}\triangleq\left\{ 1,\ldots,J\right\} $,
manage in covert wiretapping\footnote{In this paper we have focused on non-colluding $Eve$'s who try to
maximize their own SINR individually. The problem of colluding $Eve$s
has been left for future work. }. Next, the channel models and the clustering procedure operating
under CSI error are described.

\subsection{Channel Definitions }

In our scenario, the $U_{m,k}\longrightarrow E_{e,j}$ channels are
represented by $\mathbf{q}_{m,j,k},\ \forall m,\ k,\ j$, while the
legitimate channels spanning from the terrestrial user to the UAV-BS,
i.e., $U_{m,k}\longrightarrow UAV$, are denoted by $\mathbf{h}_{m,k}\ \forall m,k$.
The ground-to-air (G2A) channels are modeled by $\mathbf{h}_{m,k}=\sqrt{PL\left(d_{m,k}\right)}\,\mathbf{f}_{m,k}$,
where $PL\left(d_{m,k}\right)\triangleq{d_{m,k}}^{-\alpha_{m,k}}$
represents the large-scale fading, while $d_{m,k}$ and $\mathbf{f}_{m,k}\sim\mathcal{CN}\left(0,\mathbf{I}_{N_{t}}\right)$
therein, respectively, denote the G2A distance and the corresponding
small-scale fading. The path-loss exponent $\alpha_{m,k}$ obeys the
probabilistic model \cite{=00005B19=00005D}, which is appropriate
for low-altitude UAVs comprised of both the LoS and non-LoS components
$\mathcal{L}_{m,k}$ and $\mathcal{N}_{m,k}$, given by: 
\begin{equation}
\alpha_{m,k}\triangleq\frac{\mathcal{L}_{m,k}-\mathcal{N}_{m,k}}{1+\lambda_{1}.\exp{\left[\lambda_{2}\left(\theta_{m,k}-\lambda_{1}\right)\right]}}+\mathcal{N}_{m,k},\label{eq:1}
\end{equation}
 where $\theta_{m,k}$ denotes the elevation angle between the UAV
and user $U_{m,k}$, while $\lambda_{1}$ and $\lambda_{2}$ are the
constants determined by the propagation environment \cite{=00005B19=00005D}. 

\subsection{Clustering Under CSI Error }

\begin{figure}[tbh]
\centering{}\includegraphics[viewport=20bp 160bp 570bp 650bp,clip,scale=0.47]{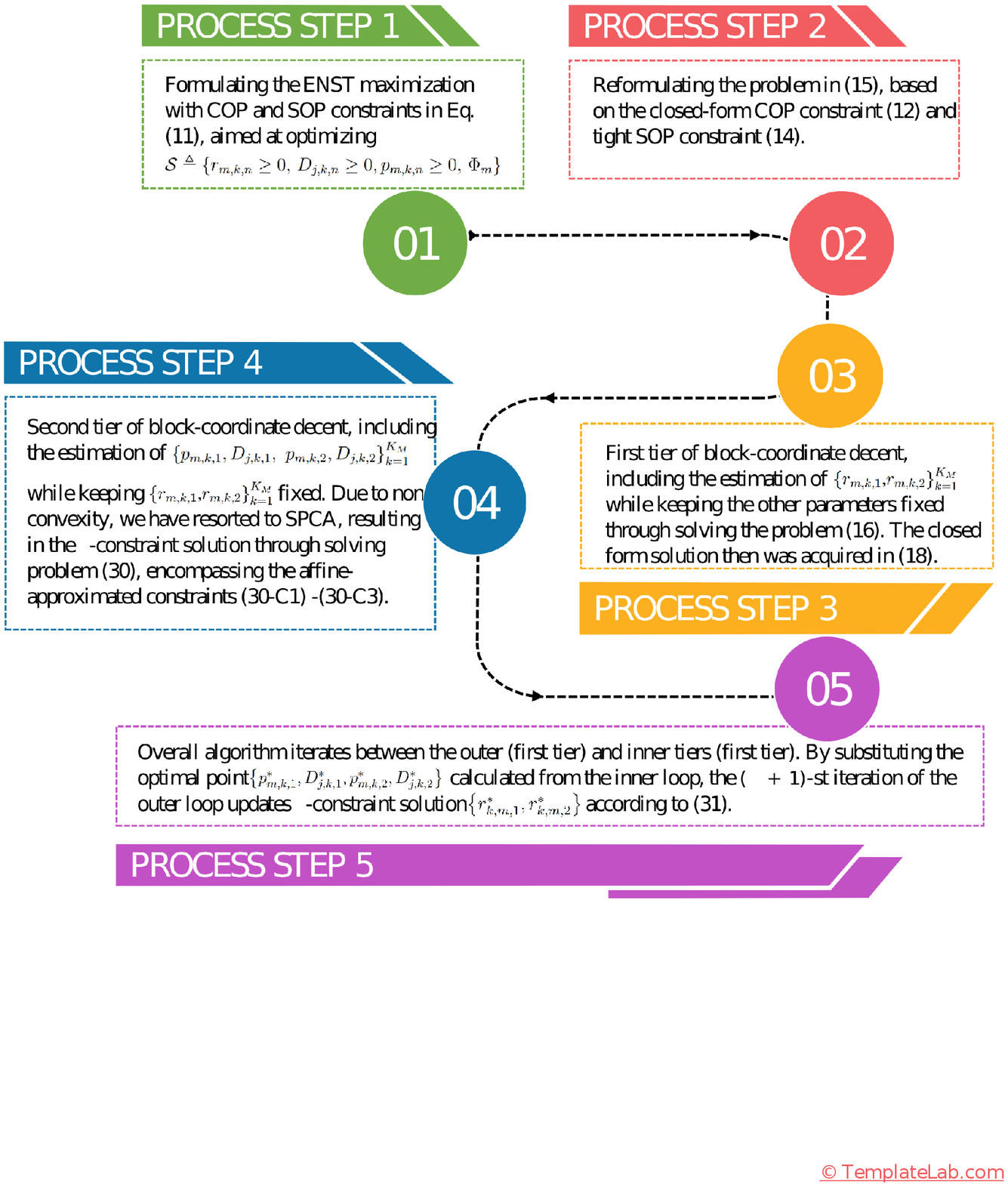}\caption{\label{fig:The-algorithmic-procedure}The algorithmic procedure of
the proposed method}
\end{figure}

Given the slowly time-varying nature of the $PL\left(d_{m,k}\right)$,
we assume that both the UAV as well as the users can estimate it perfectly.
However, due to the limited hardware complexity of the UAV, we assume
that the UAV only captures the angle-of-arrival (AoA) information
of the user-UAV channel, and even this AoA information is imperfect.
To elaborate a little further, first the UAV broadcasts a sequence
of training symbols towards the ground users, who aim for acquiring
the knowledge of their own DL channels. In general, given a sufficiently
high transmit power, as well as a long training sequence, legitimate
users are capable of perfectly estimating their own channels. More
explicitly, all the users within the $m^{th}$ cluster have the same
AoA and thus we can construct a codebook $\mathcal{V}$, comprised
of $M$ unit-norm vectors $\left\{ \mathbf{v}_{m}\right\} _{m=1}^{M}\in\mathbb{C}^{N_{t}\times1}$.
At the network's initialization, this codebook is randomly generated
and made known off-line to both the UAV and the users for example
via the codebook distribution regime of \cite{=00005B20=00005D}.
To convey the corresponding AoA to the UAV, each $U_{m,k}$ quantizes
its channel direction, i.e., $\widetilde{\mathbf{f}}_{m,k}\triangleq\frac{\mathbf{f}_{m,k}}{\left\Vert \mathbf{f}_{m,k}\right\Vert }$,
to the closest vector in terms of the chordal distance metric of \footnote{The optimal vector quantization strategy in multi-user uplink channels,
even in single-cell systems, is not known in general and is beyond
the scope of our work.} \cite{=00005B21=00005D,=00005B22=00005D}:{\small{}
\begin{align}
\hat{\mathbf{f}}_{m,k} & \triangleq\underset{\mathbf{v}_{m}\in\mathcal{V}}{\arg\max}\left|\widetilde{\mathbf{f}}_{m,k}^{H}\mathbf{v}_{m}\right|^{2}=\underset{\mathbf{v}_{m}\in\mathcal{V}}{\arg\max}\,\,\cos^{2}\left[\angle\left(\widetilde{\mathbf{f}}_{m,k},\mathbf{v}_{m}\right)\right].\label{eq:2}
\end{align}
}Accordingly, the users having the maximum chordal distance between
their so-obtained channel direction ${\widetilde{\mathbf{f}}}_{m,k}$
and $\mathbf{v}_{m}$ are allocated to the $m^{th}$ cluster and the
number of users within each cluster, i.e., $K_{m}$, is also updated
after the grouping. Now, each user sends the corresponding codebook
index back to the UAV using $B\triangleq\left\lceil \log_{2}{M}\right\rceil $
bits through an error-free and delay-free feedback channel. However,
because of the limited feedback per channel coherence block as well
as the instability of the UAV platform, the CSI of the main channel
obtained at the UAV is imperfect. Thus, a quantization error in the
form of 
\begin{equation}
\widetilde{\mathbf{f}}_{m,k}=\cos\left(\phi_{m,k}\right)\hat{\mathbf{f}}_{m,k}+\sin\left(\phi_{m,k}\right)\mathbf{e}_{m,k},\label{eq:3-1}
\end{equation}
appears in the AoA estimates of users, where $\mathbf{e}_{m,k}\in\mathbb{C}^{N_{t}\text{\texttimes}1}$
is the unit-norm quantization error vector isotropically distributed
in the null-space of $\widetilde{\mathbf{f}}_{m,k}$, while $\phi_{m,k}\triangleq\angle\left(\widetilde{\mathbf{f}}_{m,k},\mathbf{v}_{m}\right)$
of \eqref{eq:3-1} represents the angle between $\widetilde{\mathbf{f}}_{m,k}$
and $\mathbf{v}_{m}$, and $\sin^{2}\left(\phi_{m,k}\right)$ being
a random variable, whose variance is determined by $B$ \cite{=00005B23=00005D}.
To visualize the proposed approach, the whole procedure is shown in
Fig. \ref{fig:The-algorithmic-procedure}, while further details will
be presented in the sequel.

\section{Signal Representation and Problem Formulation}

In the RSMA uplink, each $U_{m,k}$ within the $m^{th}$ cluster transmits
a superposition code of two normalized sub-messages $\left.s_{m,k,n}\right|_{n=1}^{2}$,
i.e., $\mathbb{E}\left\{ \left|s_{m,k,n}\right|^{2}\right\} =1$ ,
given by \cite{=00005B24=00005D}:
\begin{equation}
s_{m,k}=\stackrel[n=1]{2}{\sum}\sqrt{p_{m,k,n}}s_{m,k,n},\,\,\forall k\text{\ensuremath{\in}}\mathcal{K}_{m},\label{eq:3}
\end{equation}
where $p_{m,k,n},\,\forall n\in\left\{ 1,2\right\} $ corresponds
to the transmit power of $s_{m,k,n}\,\forall n\in\left\{ 1,2\right\} $.
During the uplink signal reception, the UAV relies on beamforming
for discriminating the signals received by suppressing the IUI. 
\begin{figure}[tbh]
\begin{onehalfspace}
\centering{}\includegraphics[viewport=2bp 2bp 648bp 522bp,clip,scale=0.39]{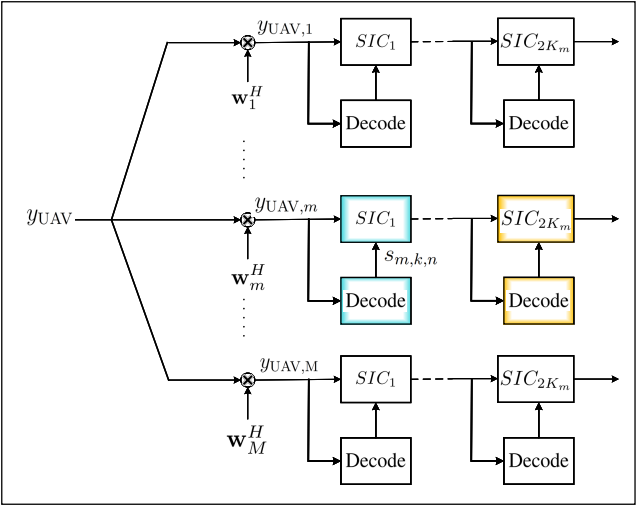}\caption{\label{fig:RSMA-empowered-BS-Structure}RSMA-based BS structure in
$M$-Cluster UL communications.}
\end{onehalfspace}
\end{figure}
Therefore, as shown in Fig. \ref{fig:RSMA-empowered-BS-Structure},
the signal received at the UAV, i.e., $y_{\mathrm{UAV}}$, is passed
through $M$ different angularly selective filters $\left\{ \mathbf{w}_{m}\right\} _{m=1}^{M}\in\mathbb{C}^{N_{t}\times1}$,
distributed in $M$ branches. Accordingly, the $m^{th}$ signal (i.e.,
the $m^{th}$ branch) received by the UAV and the received signal
at $Eves$, respectively denoted by $y_{UAV,m}$ and $y_{e,j}$, are
formulated as follows:
\begin{align}
y_{\mathrm{UAV}} & =\stackrel[i=1]{M}{\sum}\left(\stackrel[k=1]{K_{i}}{\sum}\mathbf{h}_{i,k}s_{i,k}\right)+\mathbf{z},\label{eq:4}\\
y_{\mathrm{UAV},m} & =\mathbf{w}_{m}^{H}\left(y_{\mathrm{UAV}}\right),\nonumber \\
y_{e,j} & =\mathbf{w}_{m,j}^{H}\left(\mathbf{Q}_{m,j}\mathbf{s}_{m}+\mathbf{z}_{m,j}\right),\,\forall j\text{\ensuremath{\in}}\mathcal{J},\label{eq:5}
\end{align}
where $\mathbf{Q}_{m,j}\triangleq\left[\mathbf{q}_{m,j,1},\mathbf{q}_{m,j,2},...,\mathbf{q}_{m,j,K}\right]$,
$\mathbf{s}_{m}\triangleq\left[s_{m,1},s_{m,2},...,s_{m,K}\right]^{T}$,
$z_{m}\triangleq\mathbf{w}_{m}^{H}\mathbf{z}_{m}\sim\mathcal{CN}\left(0,\sigma_{m}^{2}\right)$
and $\mathbf{z}_{m,j}\sim\mathcal{CN}\left(0,\sigma_{e}^{2}\mathbf{I}_{N_{e}}\right)$
represent the additive white Gaussian noise (AWGN) due to the $m^{th}$
cluster at the UAV and at the $j^{th}$ $Eve$, respectively. ${\color{red}{\color{black}\mathbf{w}_{m,j}\triangleq\mathbf{u}^{max}\left\{ \tilde{\mathcal{\mathbf{Q}}}_{m,j}\right\} }}$
represents the MRC beamformer employed by $Eve$ where $\tilde{\mathcal{\mathbf{Q}}}_{m,j}\triangleq\mathbf{Q}_{m,j}\left(\mathbf{Q}_{m,j}\right)^{H}$.
It is easy to show that $y_{e,j}=\stackrel[k=1]{K_{m}}{\sum}g_{m,k,j}s_{m,k}+\xi_{e,j},\,\forall j\text{\ensuremath{\in}}\mathcal{J},$
where $g_{m,k,j}\sim\mathcal{CN}\left(0,\sqrt{\lambda_{j}^{max}\left\{ \tilde{\mathcal{\mathbf{Q}}}_{m,j}\right\} }\right)$
\cite{19_}, and $\xi_{e,j}\sim\mathcal{CN}\left(0,\sigma_{e}^{2}\right)$.
In terms of the worst-case secrecy scenario, $Eve$ is assumed to
be able to perfectly estimate its corresponding CSI and no ICI is
available to degrade the performance of $Eve$. To force the ICI terms
to zero, we resort the zero-forcing (ZF) beamforming. More explicitly,
upon relying on the codebook $\mathcal{V}$ discussed earlier in Section
$\textrm{I.B}$, $\mathbf{w}_{m}$ is chosen so that we have $\mathbf{w}_{m}^{H}\mathbf{v}_{l}=0,\,\forall l\neq m,\,l\text{\ensuremath{\in}}\left\{ 1,2,...,M\right\} $.

when considering the signal extracted from the $m^{th}$ cluster,
the UAV employs $2K_{m}$ number of SIC stages to suppress IUI as
well as to decode all transmitted messages in the set $\mathcal{K}_{m,n}\triangleq\left\{ s_{m,k,n}\right\} $
received from $y_{\mathrm{UAV},m}$, as illustrated in Fig. \ref{fig:RSMA-empowered-BS-Structure}.
The decoding order of the $m^{th}$ cluster at the UAV is denoted
by a permutation $\boldsymbol{\mathbf{\Phi}}_{m}$, which belongs
to set $\mathbf{\Pi}_{m}$ defined as the set of all possible decoding
orders of all $2K_{m}$ messages arriving from $K_{m}$ users, which
includes $\frac{2K_{m}!}{2^{K_{m}}}$ elements. Let $\mathrm{\Phi}_{m,k,n}$
represents the position of the message $s_{m,k,n}$ in $\boldsymbol{\mathbf{\Phi}}_{m}$.
Therefore, we can define $\Phi_{m,k,n}=\left\{ \left.\left(k',n'\right)\neq\left(k,n\right)\right|\left(k^{\prime},n^{\prime}\right)\succ\left(k,n\right)\right\} $,
where the operator $\left(k^{\prime},n^{\prime}\right)\succ\left(k,n\right)$
indicates that $s_{m,k,n}$ has a higher decoding order than $s_{m,k^{\prime},n^{\prime}}$
in $\boldsymbol{\mathbf{\Phi}}_{m}$ , i.e., the UAV is scheduled
to decode $s_{m,k',n'}$ after decoding and cancelling out the effect
of $s_{m,k,n}$. 

Therefore, in the SIC scheme of the RSMA uplink, the UAV first decodes
and subtracts the remodulated signals having higher decoding orders.
i.e., $\left(k^{\prime},n^{\prime}\right)\in\mathcal{K}_{m,n}\setminus\left\{ {{\left(k,n\right)}\bigcup\Phi_{m,k,n}}\right\} $,
then it decodes signal $s_{m,k,n}$, where the signal of the users
in $\Phi_{m,k,n}$ is treated as noise. According to the SIC protocol,
the signal of $s_{m,k,n}$ will be decoded prior to $s_{m,k',n'}$
if we have $\left|\Phi_{m,k,n}\right|>\left|\Phi_{m,k',n'}\right|$,
where $|\,\mathcal{A}\,|$ is the cardinality of the set $\mathcal{A}$.
Accordingly, the signal-to-interference-plus-noise ratio (SINR) at
the UAV experienced upon detecting $s_{m,k,n}$, and denoted by $\rho_{m,k,n}$
is formulated as \eqref{eq:6},
\begin{figure*}[tbh]
{\small{}
\begin{align}
\rho_{m,k,n} & =\frac{p_{m,k,n}\left|\mathbf{w}_{m}^{H}\mathbf{h}_{m,k}\right|^{2}}{\underset{_{\left(k',n'\right)\in\Phi_{m,k,n}}}{\sum}p_{m,k',n'}\left|\mathbf{w}_{m}^{H}\mathbf{h}_{m,k'}\right|^{2}+\stackrel[i=1,i\neq m]{M}{\sum}\,\stackrel[k''=1]{K_{i}}{\sum}P_{i,k''}\left|\mathbf{w}_{m}^{H}\mathbf{h}_{i,k''}\right|^{2}+\sigma_{m}^{2}}\label{eq:6}
\end{align}
\[
=\frac{p_{m,k,n}\mathrm{PL}\left(d_{m,k}\right)\left|\mathbf{w}_{m}^{H}\mathbf{f}_{m,k}\right|^{2}}{\underset{\mathrm{IUI}}{\underbrace{\underset{_{\left(k',n'\right)\in\Phi_{m,k,n}}}{\sum}p_{m,k',n'}\mathrm{PL}\left(d_{m,k'}\right)\,\left|\mathbf{w}_{m}^{H}\mathbf{f}_{m,k'}\right|^{2}}}+\underset{\mathrm{ICI}}{\underbrace{\stackrel[i=1,i\neq m]{M}{\sum}\,\stackrel[k''=1]{K_{i}}{\sum}P_{i,k''}\mathrm{PL}\left(d_{i,k''}\right)\left\Vert \sin\left(\phi_{i,k''}\right)\mathbf{f}_{i,k''}\right\Vert ^{2}\left|\mathbf{w}_{m}^{H}\mathbf{e}_{i,k''}\right|^{2}}}+\sigma_{m}^{2}}
\]
\_\_\_\_\_\_\_\_\_\_\_\_\_\_\_\_\_\_\_\_\_\_\_\_\_\_\_\_\_\_\_\_\_\_\_\_\_\_\_\_\_\_\_\_\_\_\_\_\_\_\_\_\_\_\_\_\_\_\_\_\_\_\_\_\_\_\_\_\_\_\_\_\_\_\_\_\_\_\_\_\_\_\_\_\_\_\_\_\_\_\_\_\_\_\_\_\_\_\_\_\_\_}
\end{figure*}
where the IUI and ICI terms are obtained by substituting (\ref{eq:3})
into (\ref{eq:6}). Notably, after clustering in the presence of beamforming
weight quantization errors the ICI cannot be completely removed by
the beamformer having the weights of $\mathbf{w}_{m}$, thus a residual
ICI term contaminates the corresponding received signal of the $m^{th}$
cluster. In other words, the beamformer weights $\mathbf{w}_{m}$
fail to perfectly null out the ICI due to the limited feedback. 

Furthermore, it is assumed that the $Eves$ have no information about
$\boldsymbol{\mathbf{\Phi}}_{m}$, hence they cannot perform SIC within
a cluster \footnote{As a more challenging secrecy scenario, for comparison, in our simulation
we consider a scenario when $E_{e,j}$ can exploit the optimal SIC
decoding order and receives no CSI.}. Consequently, from the perspective of $E_{e,j}$, the received SINR
of decoding $s_{m,k,n}$, while treating the other ones as noise,
is formulated as: {\small{}
\begin{equation}
\mu_{j,k,n}=\frac{p_{m,k,n}\mathrm{PL}\left(d_{m,k,j}\right)\left|g_{m,k,j}\right|^{2}}{\sum_{\left(k',n'\right)\neq\left(k,n\right)}p_{m,k',n'}\mathrm{PL}\left(d_{m,k',j}\right)\left|g_{m,k',j}\right|^{2}+\sigma_{e}^{2}}.\label{eq:7}
\end{equation}
}{\small\par}

Given the SINRs in \eqref{eq:6} and \eqref{eq:7}, the corresponding
achievable rates are respectively given by $C_{m,k,n}\triangleq\log_{2}{\left(1+\rho_{k,m,n}\right)}$
and $C_{j,k,n}\triangleq\log_{2}{\left(1+\mu_{j,m,n}\right)}$. 

\noindent \textit{Remark 1.} For ensuring reliable uplink communication,
the transmission rate $r_{m,k,n}$ of each $U_{m,k}$ should not exceed
$C_{m,k,n}$, i.e., $C_{m,k,n}\geq r_{m,k,n}$. However, as a consequence
of ICI and fading, $r_{m,k,n}$ might violate this condition, hence
leading to the link outages. However, the COP, defined as the probability
that a system is unable to support the target transmission rate $C_{m,k,n}$,
must be limited by the maximum tolerable COP $\epsilon_{cop}\in\left(0,1\right)$,
as follows: 
\begin{equation}
\textrm{COP}:\qquad P_{m,k,n}^{CO}\triangleq\mathbb{P}\left\{ r_{m,k,n}>C_{m,k,n}\right\} \leq\epsilon_{cop}.\label{eq:8}
\end{equation}

\noindent \textit{Remark 2.} On the other hand, since the $U_{m,k}$
has no knowledge concerning the CSIs of passive $Eves$ \cite{25_1},
the values of $\mu_{j,m,n}$ are unknown. A beneficial secrecy policy
in this situation is to adjust the redundancy rate of $U_{m,k}$ \cite{25_1},
denoted by $D_{j,k,n}$, so that the COP limit of \eqref{eq:8} is
satisfied. In other words, $D_{j,k,n}$ must not exceed $C_{j,k,n}$.
To do so, the SOP of $\epsilon_{sop}\in\left(0,1\right)$, satisfies:
\footnote{It should be highlighted that, we considered the worst-case condition
for ensuring both reliability and security of each part of split messages.
This implies that if actual transmission rate of each split messages
$s_{m,k,n}\,\forall n\in\left\{ 1,2\right\} $ corresponding to $U_{m,k}$
can satisfy its target transmission rate $C_{m,k,n}\,\forall n\in\left\{ 1,2\right\} $,
as stipulated in the COP condition of \eqref{eq:8}, we can guarantee
that the per user basis condition is also met. At the receiver, UAV-BS
recover and merges each of these two split messages corresponding
to each user separately to retrieve their original messages. Thus,
the conditions described above, should be satisfied for each part
of message separately. We can use the same justification for the SOP
constraint \eqref{eq:9}.}
\begin{equation}
\textrm{SOP}:\qquad P_{j,k,n}^{SO}\triangleq\mathbb{P}\left\{ D_{j,k,n}\leq C_{j,k,n}\right\} \leq\epsilon_{sop}.\label{eq:9}
\end{equation}

\noindent \textit{Remark 3.} Upon considering non-colluding $Eves$,
the achievable secrecy rates of $U_{m,k}$ where transmitting $s_{k,m,n}$
is limited by the worst-case $Eve$ scenario of $C_{m,k,n}^{sec}\triangleq$
$\underset{1\text{\ensuremath{\le}}j\text{\ensuremath{\le}}J}{\min}\left\{ \left[r_{m,k,n}\text{\textminus}D_{j,k,n}\right]^{+}\right\} $.
On the other hand, while minimizing $P_{m,k,n}^{CO}$ would improve
the reliability, maximizing $C_{m,k,n}^{sec}$ will enhance the security
upon jointly considering both the reliability and security requirements
of all $\left.U_{m,k}\right|_{k=1}^{\mathcal{K}_{m}}$, we should
rather maximize the ENST, defined as $C_{ENST}\triangleq\sum_{k=1}^{K_{m}}$$\sum_{n=1}^{2}\left(1-P_{m,k,n}^{CO}\right)\,C_{m,k,n}^{sec}$. 

Based on the discussion in Remarks 1-3, while considering the limited
power budget imposed on each $U_{m,k}$, formulated as $\sum_{n=1}^{2}p_{m,k,n}\le P_{m,k}$,
the optimization problem of the proposed secure RSMA-based uplink
is formulated as:{\small{}
\begin{equation}
\underset{\mathcal{S}}{\textrm{max}}\left(\underset{1\text{\ensuremath{\le}}j\text{\ensuremath{\le}}J}{\min}\left\{ \stackrel[k=1]{K_{m}}{\sum}\stackrel[n=1]{2}{\sum}\left(1\text{\textminus}P_{m,k,n}^{CO}\right)\left[r_{m,k,n}\text{\textminus}D_{j,k,n}\right]^{+}\right\} \right)\,\,\label{eq:10}
\end{equation}
}s.t.

\noindent {\small{}}%
\begin{tabular}{>{\raggedright}p{9cm}}
{\small{}$C_{1}:$ $P_{m,k,n}^{CO}\leq{\epsilon_{cop}},\,\,\,\forall\,k$,\vspace{0.1cm}
}\tabularnewline
\end{tabular}{\small\par}

\noindent {\small{}}%
\begin{tabular}{>{\raggedright}p{9cm}}
{\small{}$C_{2}:$ $P_{j,k,n}^{SO}\leq{\epsilon_{sop}},\,\,\forall\,k,j$,\vspace{0.1cm}
}\tabularnewline
\end{tabular}{\small\par}

\noindent {\small{}}%
\begin{tabular}{>{\raggedright}p{9cm}}
{\small{}$C_{3}:$ $\stackrel[n=1]{2}{\sum}p_{m,k,n}\leq P_{m,k},\,\,p_{m,k,n}\geq0,\,\,\forall\,k,$
\vspace{0.1cm}
}\tabularnewline
\end{tabular}{\small\par}

\noindent where $\mathcal{S}\triangleq\left\{ r_{m,k,n}\geq0,\,D_{j,k,n}\geq0,p_{m,k,n}\geq0,\,\Phi_{m}\right\} $.
Due to the non-convex OF, as well as the discontinuous variable ${\Phi}_{m}$,
the problem in \eqref{eq:10} represents a non-convex mixed integer
programming problem. In the next section, we derive closed-form expressions
both for the COP and SOP constraints, while ${\Phi}_{m}$ is obtained
through an exhaustive search. 

\section{ENST Maximization Solution}

In this section, we construct the overall algorithm for finding the
optimal solution of \eqref{eq:10}. 

\subsection{Handling the Probabilistic Constraints \eqref{eq:10}-$C_{1}$ and
\eqref{eq:10}-$C_{2}$}

We first intend to handle the COP constraint \eqref{eq:10}-$C_{1}$
. In this regard, we first insert \eqref{eq:6} into \eqref{eq:8},
so that \eqref{eq:10}-$C_{1}$ may be reformulated as \eqref{eq:11}
\begin{figure*}[tbh]
\centering{}%
\begin{tabular}{>{\raggedright}p{17.5cm}}
{\small{}
\begin{equation}
P_{m,k,n}^{CO}\triangleq\mathbb{P}\left\{ r_{m,k,n}>\log_{2}\left(1+\rho_{m,k,n}\right)\right\} \label{eq:11}
\end{equation}
}\tabularnewline
\raggedright{}{\small{}
\[
=1-\exp\left(-\frac{\beta_{m,n,k}\sigma_{m}^{2}}{2}\right)\stackrel[\left(k',n'\right)\in\Phi_{m,k,n}]{}{\prod}\left(1+\lambda_{m,k',n'}^{-1}\frac{\beta_{m,n,k}}{2}\right)^{-1}\stackrel[i=1,i\neq m]{M}{\prod}\stackrel[k''=1]{K_{i}}{\prod}\left(1+\lambda_{i,k''}^{-1}\frac{\beta_{m,n,k}}{2}\right)^{-1},\forall k\text{\ensuremath{\in}}\mathcal{K}_{m}
\]
}\tabularnewline
\bottomrule
\end{tabular}
\end{figure*}
, where we have $\beta_{m,n,k}=2^{r_{m,k,n}}-1$, $\lambda_{m,k',n'}\triangleq$$\frac{1}{2p_{m,k',n'}\mathrm{PL}\left(d_{m,k'}\right)}$,
and $\lambda_{i,k''}=$$\frac{2^{\frac{B}{N_{t}-1}}}{P_{i,k''}\mathrm{PL}\left(d_{i,k''}\right)}$
(\textit{Proof}: See Appendix A).\vspace{0.1cm}

Upon inserting \eqref{eq:7} into \eqref{eq:9}, we may reformulate
the SOP constraint \eqref{eq:10}-$C_{2}$ as \eqref{eq:12}, 
\begin{figure*}[tbh]
{\small{}
\begin{align}
P_{j,k,n}^{SO} & \triangleq\mathbb{P}\left\{ D_{j,k,n}\leq\log_{2}\left(1+\mu_{j,k,n}\right)\right\} =\exp\left(-\eta_{j,k,n}\kappa_{j,k,n}\sigma_{e}^{2}\right)\stackrel[\left(k',n'\right)\neq\left(k,n\right)]{}{\prod}\left(1+\eta_{j,k,n}\kappa_{j,k,n}\zeta_{j,k',n'}^{-1}\right)^{-1},\label{eq:12}
\end{align}
}\_\_\_\_\_\_\_\_\_\_\_\_\_\_\_\_\_\_\_\_\_\_\_\_\_\_\_\_\_\_\_\_\_\_\_\_\_\_\_\_\_\_\_\_\_\_\_\_\_\_\_\_\_\_\_\_\_\_\_\_\_\_\_\_\_\_\_\_\_\_\_\_\_\_\_\_\_\_\_\_\_\_\_\_\_\_
\end{figure*}
where we have $\eta_{j,k,n}=\frac{1}{p_{m,k,n}\mathrm{PL}\left(d_{m,k,j}\right)\lambda_{j}^{max}\left\{ \tilde{\mathcal{\mathbf{Q}}}_{m,j}\right\} }$,
$\zeta_{j,k',n'}=\frac{1}{p_{m,k',n'}\mathrm{PL}\left(d_{m,k',j}\right)\lambda_{j}^{max}\left\{ \tilde{\mathcal{\mathbf{Q}}}_{m,j}\right\} }$,
and $\kappa_{j,k,n}=2^{D_{j,k,n}}-1$ (\textit{Proof}: See Appendix
B). On the other hand, since $D_{j,k,n}$ independent of both $r_{k,m,n}$
and $p_{k,m,n}$ within the OF, the maximization problem \eqref{eq:10}
over $D_{j,k,n}$ is equivalent to minimizing $D_{j,k,n}$. To find
a more conservative solution, we exploit that $D_{j,k,n}$ appears
both in the OF and in the SOP constraint \eqref{eq:10}, we have to
exploit a tighter constraint than \eqref{eq:12} for obtaining the
minimum value of $D_{j,k,n}$, which is given by \eqref{eq:14}, 
\begin{figure*}[tbh]
\begin{tabular}{>{\raggedright}p{17.5cm}}
{\small{}
\begin{equation}
D_{j,k,n}\leq\log_{2}\left(1+\frac{2K_{m}-1}{\eta_{j,k,n}\sigma_{e}^{2}}W_{0}\left(\frac{\eta_{j,k,n}\sigma_{e}^{2}}{2K_{m}-1}\left(\frac{\stackrel[\left(k',n'\right)\neq\left(k,n\right)]{}{\prod}\zeta_{j,k',n'}}{\eta_{j,k,n}}{\epsilon_{sop}}^{-1}\right)^{\frac{1}{2K_{m}-1}}\right)\right),\label{eq:14}
\end{equation}
}\tabularnewline
\bottomrule
\end{tabular}
\end{figure*}
where $W_{0}\left(x\right)$ is the Lambert $W$-function (\textit{Proof}:
See Appendix C). However, it is still challenging to solve \eqref{eq:10},
since $r_{k,m,n}$ and $p_{k,m,n}$ are coupled in the OF of \eqref{eq:10}.
To arrive at a more tractable form, the operations of maximization
and the minimization can be swapped in \eqref{eq:10}. Additionally,
since $\left\{ D_{j,k,n}\right\} _{j=1}^{M}$ are independent, we
can actually solve $J$ independent maximization problems and then
simply choose the minimum one. Furthermore, by exploiting the inequalities
of $\exp\left(-x\right)\leq\frac{1}{1+x}$ and $\frac{1}{1+x}\leq\frac{1}{x}$,
we can instead replace the lower bound and upper bound of the OF and
of the COP constraint \eqref{eq:10}, respectively. Accordingly, based
on what was mentioned above, a bound of the solution may be obtained
as \eqref{eq:15}, 
\begin{figure*}[tbh]
\begin{equation}
\underset{1\text{\ensuremath{\le}}j\text{\ensuremath{\le}}J}{\min}\left(\underset{\mathcal{S}}{\max}\left\{ \stackrel[k=1]{K_{m}}{\sum}\stackrel[n=1]{2}{\sum}\exp\left(\frac{\beta_{m,n,k}}{2}\left[\stackrel[\left(k',n'\right)\in\Phi_{m,k,n}]{}{\sum}\lambda_{m,k',n'}^{-1}+\xi-\sigma_{m}^{2}\right]\right)\left[r_{m,k,n}\text{\textminus}D_{j,k,n}\right]^{+}\right\} \right)\,\,\label{eq:15}
\end{equation}
s.t.

\begin{tabular}{>{\raggedright}p{15cm}}
$C_{1}:$ $\xi\exp\left(-\frac{\beta_{m,n,k}\sigma_{m}^{2}}{2}\right)\beta_{m,n,k}^{-A}\stackrel[\left(k',n'\right)\in\Phi_{m,k,n}]{}{\prod}2\lambda_{m,k',n'}^{-1}\leq{\epsilon_{cop}},\,\,\,\forall\,k,n$\vspace{0.1cm}
\tabularnewline
\end{tabular}

\begin{tabular}{>{\raggedright}p{16cm}}
$C_{2}:$ $D_{j,k,n}\leq\log_{2}\left(1+\frac{2K_{m}-1}{\eta_{j,k,n}\sigma_{e}^{2}}W_{0}\left(\frac{\eta_{j,k,n}\sigma_{e}^{2}}{2K_{m}-1}\left(\frac{\stackrel[\left(k',n'\right)\neq\left(k,n\right)]{}{\prod}\zeta_{j,k',n'}}{\eta_{j,k,n}}{\epsilon_{sop}}^{-1}\right)^{\frac{1}{2K_{m}-1}}\right)\right),\,\,\forall\,k,j,n$\vspace{0.1cm}
\tabularnewline
\end{tabular}

\begin{tabular}{>{\raggedright}p{17.7cm}}
$C_{3}:$ $\stackrel[n=1]{2}{\sum}p_{m,k,n}\leq P_{m,k},\,\,p_{m,k,n}\geq0,\,\,\forall\,k,$
\vspace{0.1cm}
\tabularnewline
\hline 
\end{tabular}
\end{figure*}
where we have $A\triangleq\left|\Phi_{m,k,n}\right|$$+M+K_{i}$,
$\xi\triangleq\stackrel[i=1,i\neq m]{M}{\prod}\stackrel[k''=1]{K_{i}}{\prod}2\lambda_{i,k''}^{-1}$.

Now, we can exploit the block coordinate decent technique, where $\left\{ p_{m,k,1},D_{j,k,1},\ p_{m,k,2},D_{j,k,2}\right\} _{k=1}^{K_{M}}$
and the $\left\{ r_{m,k,1}\text{,\ensuremath{r_{m,k,2}}}\right\} _{k=1}^{K_{M}}$
are found successively in an iterative manner. In particular, the
$l^{th}$ iteration of the algorithm is constituted by separately
maximizing the criterion with respect to each of $\left\{ r_{m,k,1}\text{,\ensuremath{r_{m,k,2}}}\right\} _{k=1}^{K_{M}}$
and $\left\{ p_{m,k,1},D_{j,k,1},\ p_{m,k,2},D_{j,k,2}\right\} _{k=1}^{K_{M}}$,
while keeping the other one fixed. Given this perspective, we first
update $\left\{ r_{m,k,1}\text{,\ensuremath{r_{m,k,2}}}\right\} _{k=1}^{K_{M}}$,
while assuming that $\left\{ p_{m,k,1},D_{j,k,1},\ p_{m,k,2},D_{j,k,2}\right\} _{k=1}^{K_{M}}$
are fixed values which results in the following optimization problem:\vspace{0.1cm}
{\small{}
\begin{equation}
\underset{1\text{\ensuremath{\le}}j\text{\ensuremath{\le}}J}{\min}\left(\underset{\left\{ r_{m,k,1},r_{m,k,2}\right\} }{\max}\left\{ \stackrel[k=1]{K_{m}}{\sum}\stackrel[n=1]{2}{\sum}\exp\left(\Xi_{m,n}^{r}\frac{\beta_{m,n,k}}{2}\right)r_{m,k,n}\right\} \right),\,\,\label{eq:16}
\end{equation}
}s.t.

\noindent %
\begin{tabular}{>{\raggedright}p{10cm}}
{\small{}$\xi\exp\left(-\frac{\beta_{m,n,k}\sigma_{m}^{2}}{2}\right)\beta_{m,n,k}^{-A}\stackrel[\left(k',n'\right)\in\Phi_{m,k,n}]{}{\prod}\left(2\lambda_{m,k',n'}^{*^{-1}}\right)\leq{\epsilon_{cop}},$\vspace{0.1cm}
}\tabularnewline
\end{tabular}

\noindent $\forall\,k,n,$ where $\Xi_{m,n}^{r}\triangleq$$\stackrel[\left(k',n'\right)\in\Phi_{m,k,n}]{}{\sum}\lambda_{m,k',n'}^{*^{-1}}$$+\xi-\sigma_{m}^{2}$.

It is easy to check that in terms of $\left\{ r_{m,k,1}\text{,\ensuremath{r_{m,k,2}}}\right\} _{k=1}^{K_{M}}$,
while the OF of \eqref{eq:16} is an increasing function, while constraint
\eqref{eq:16}-$C_{1}$ is a decreasing one. Hence, the closed form
expression of $\left\{ r_{m,k,1}\text{,\ensuremath{r_{m,k,2}}}\right\} _{k=1}^{K_{M}}$
may obtained when the inequality constraint \eqref{eq:16}-$C_{1}$
is active at the optimum. Hence, the optimal point of (\ref{eq:16}),
i.e., $\left\{ r_{m,k,1}^{*}\text{,\ensuremath{r_{m,k,2}}}^{*}\right\} _{k=1}^{K_{M}}$,
may be found by solving the following equation:{\small{}
\begin{equation}
\exp\left(-\frac{\sigma_{m}^{2}\beta_{m,n,k}}{2}\right)\beta_{m,n,k}^{-A}=\frac{{\epsilon_{cop}}}{\xi}\stackrel[\left(k',n'\right)\in\Phi_{m,k,n}]{}{\prod}\left(\frac{\lambda_{m,k',n'}^{*}}{2}\right).\label{eq:17}
\end{equation}
}{\small\par}

By doing so, followed by some algebraic manipulations, together with
the help of the principal branch of the Lambert $W$-function, the
optimum is formulated as follows:{\footnotesize{}
\begin{equation}
r_{m,k,n}^{*}=\log_{2}\left(1+\frac{2A}{\sigma_{m}^{2}}W_{0}\left(\left[\xi{\epsilon_{cop}}^{-1}\stackrel[\left(k',n'\right)\in\Phi_{m,k,n}]{}{\prod}\left(2\lambda_{m,k',n'}^{*^{-1}}\right)\right]^{\frac{1}{A}}\right)\right).\label{eq:18}
\end{equation}
}{\footnotesize\par}

\begin{figure*}[tbh]
\begin{equation}
\underset{1\text{\ensuremath{\le}}j\text{\ensuremath{\le}}J}{\min}\left(\underset{\left\{ p_{m,k,1},D_{j,k,1},\ p_{m,k,2},D_{j,k,2}\right\} _{k=1}^{K_{M}}}{\max}\left\{ \mathcal{F}_{j}\right\} \right),\,\,\label{eq:19}
\end{equation}
s.t.

\begin{tabular}{>{\raggedright}p{7cm}}
$C_{1}:$ $\stackrel[\left(k',n'\right)\in\Phi_{m,k,n}]{}{\sum}\log_{2}\left(p_{m,k',n'}\right)\leq\psi_{k},$\vspace{0.1cm}
\tabularnewline
\end{tabular}%
\begin{tabular}{>{\raggedright}p{4cm}}
$C_{3}:$ $r_{m,k,n}^{*}\geq D_{j,k,n},$\vspace{0.1cm}
\tabularnewline
\end{tabular}%
\begin{tabular}{>{\raggedright}p{9cm}}
$C_{4}:$ $\stackrel[n=1]{2}{\sum}p_{m,k,n}\leq P_{m,k},\,p_{m,k,n}\geq0,$
\vspace{0.1cm}
\tabularnewline
\end{tabular}

\begin{tabular}{>{\raggedright}p{17.5cm}}
$C_{2}:$ $D_{j,k,n}\leq\log_{2}\left(1+\frac{2K_{m}-1}{\eta_{j,k,n}\sigma_{e}^{2}}W_{0}\left(\frac{\eta_{j,k,n}\sigma_{e}^{2}}{2K_{m}-1}\left(\frac{\stackrel[\left(k',n'\right)\neq\left(k,n\right)]{}{\prod}\zeta_{j,k',n'}}{\eta_{j,k,n}}{\epsilon_{sop}}^{-1}\right)^{\frac{1}{2K_{m}-1}}\right)\right),\,\,\forall\,k,j,n$\vspace{0.1cm}
\tabularnewline
\hline 
\end{tabular}
\end{figure*}
Now, assuming $\left\{ r_{m,k,1}^{*}\text{,\ensuremath{r_{m,k,2}}}^{*}\right\} _{k=1}^{K_{m}}$
to be fixed values, we can update $\left\{ p_{m,k,1},D_{j,k,1},\ p_{m,k,2},D_{j,k,2}\right\} _{k=1}^{K_{M}}$,
which result in the following equivalent transformation of (\ref{eq:16})
in the log-domain as \eqref{eq:19}, where $\mathcal{F}_{j}$ $\triangleq$
$\stackrel[k=1]{K_{m}}{\sum}\stackrel[n=1]{2}{\sum}\beta_{m,n,k}^{*}$
$\left[\stackrel[\left(k',n'\right)\in\Phi_{m,k,n}]{}{\sum}\mathrm{PL}\left(d_{m,k'}\right)p_{m,k',n'}\right]$$+\ln\left[r_{m,k,n}^{*}\text{\textminus}D_{j,k,n}\right]$,
$\gamma\triangleq2^{-2\left|\Phi_{m,k,n}\right|}$ ${\epsilon_{cop}}\xi^{-1}$
$\exp\left(\frac{\beta_{m,n,k}^{*}\sigma_{m}^{2}}{2}\right)$ $\beta_{m,n,k}^{*^{\left(\left|\Phi_{m,k,n}\right|+M+K_{i}\right)}}$,
and $\psi_{k}\triangleq\log_{2}\left({\gamma}\right)-$$\stackrel[\left(k',n'\right)\in\Phi_{m,k,n}]{}{\sum}\log_{2}\left(\mathrm{PL}\left(d_{m,k'}\right)\right)$.

Note that the newly added constraint \eqref{eq:19}-$C_{3}$ arises
from the fact that the point-wise maximum operator $\left[\,\right]^{+}$
within the OF of \eqref{eq:15} leads to non-convexity. Thus, by adding
\eqref{eq:19}-$C_{3}$ we are equivalently stating that the OF must
be non-negative at the optimum and then we can simply remove $\left[\,\right]^{+}$.
Now, since the OF in \eqref{eq:19} is constituted by the sum of convex
and affine functions with respect to $\left\{ p_{m,k,1},D_{j,k,1},\ p_{m,k,2},D_{j,k,2}\right\} _{k=1}^{K_{M}}$,
it is a convex function. However, problem \eqref{eq:19} is still
non-convex because of the non-convex constraint \eqref{eq:19}-$C_{2}$
. To circumvent the non-convexity, we harness the SPCA of \cite{=00005B4=00005D,=00005B14=00005D},
where the non-convex factor is approximated by its first-order Taylor
expansion at each iteration. Given this perspective, in the following
we attempt to circumvent the non-convexity imposed by the fractional
form and the logarithmic function within \eqref{eq:19}-$C_{2}$ by
introducing some auxiliary variables, namely $\left\{ \theta_{j,k,n},\varrho_{j,k,n},\nu_{j,k,n},\vartheta_{j,k,n}\right\} $.
Following the classic variable transformation approach, the constraint
\eqref{eq:19}-$C_{2}$ can be decomposed into $\forall\,k,j,n,$:{\small{}}
\begin{algorithm}[tbh]
{\small{}\caption{\textbf{: Secure resource allocation proposed for the RSMA uplink}}
}{\small\par}

{\small{}For ${\Phi}_{m}$ $\in$ $\mathbf{\Pi}_{m}$ do:}{\small\par}

{\small{}~~~~~~~Call }\textbf{\small{}Function}\textit{\small{}
Outer\_Loop}{\small\par}

{\small{}End.}{\small\par}

{\small{}Obtain the optimal solution$\biggl\{ p_{m,k,1}^{*},D_{j,k,1}^{*},p_{m,k,2}^{*}$
$,D_{j,k,2}^{*}\biggr\}_{k=1}^{K_{m}}$ , $\left\{ r_{m,k,1}^{*},r_{m,k,2}^{*}\right\} _{k=1}^{K_{m}}$
and optimal decoding order ${\Phi}_{m}^{*}$=${\Phi}_{m}$ with the
highest OF.}{\small\par}

{\small{}=============================================}{\small\par}

\textbf{\small{}Function}{\small{} }\textit{\small{}Outer\_Loop}{\small\par}

{\small{}Step 1: Initialize the maximum number of iterations $Q_{max}$,
$T_{max}$ and the maximum tolerance $\epsilon$.}{\small\par}

{\small{}Step 2: Initialize $\left\{ r_{m,k,1}^{*}{}^{\left[0\right]},r_{m,k,2}^{*}{}^{\left[0\right]}\right\} $
and the outer iteration index $q=0$.}{\small\par}

{\small{}While $\left(\left|r_{m,k,n}^{*}{}^{\left[q+1\right]}-r_{m,k,n}^{*}{}^{\left[q\right]}\right|\geq\epsilon\,\textrm{or}\,q\le Q_{max}\right)$
$\forall\,k,n$, do:\vspace{0.05cm}
}{\small\par}

{\small{}~~}%
\begin{tabular}{|>{\raggedright}p{8.6cm}}
{\small{}Step 3: Call the }\textbf{\small{}Function}{\small{} }\textit{\small{}Inner\_Loop}{\small{}
with $\left\{ r_{m,k,1}^{*}{}^{\left[q\right]},r_{m,k,2}^{*}{}^{\left[q\right]}\right\} $
to obtain the $\epsilon$-constraint solution $\left\{ p_{m,k,1}^{*},D_{j,k,1}^{*},p_{m,k,2}^{*},D_{j,k,2}^{*}\right\} $.}{\small\par}

{\small{}Step 4: Update $r_{m,k,n}^{*}{}^{\left[q+1\right]}$ in \eqref{eq:18}.}{\small\par}

{\small{}Step 5: }\textbf{\small{}Goto}{\small{} Step 3.}\tabularnewline
\end{tabular}{\small{}\vspace{0.05cm}
}{\small\par}

{\small{}end while.}{\small\par}

{\small{}Step 6: Return the $\epsilon$-constraint solution $\left\{ p_{m,k,1}^{*},D_{j,k,1}^{*},p_{m,k,2}^{*},D_{j,k,2}^{*}\right\} $,
$r_{m,k,1}^{*}=r_{m,k,1}^{*}{}^{\left[q+1\right]}$ and $r_{m,k,2}^{*}=r_{m,k,2}^{*}{}^{\left[q+1\right]}$
.}{\small\par}

{\small{}end.}{\small\par}

{\small{}=============================================}{\small\par}

\textbf{\small{}Function}{\small{} }\textit{\small{}Inner\_Loop}{\small{}
$\left(\left\{ r_{m,k,1}^{*}{}^{\left[q+1\right]},r_{m,k,2}^{*}{}^{\left[q+1\right]}\right\} \right)$}{\small\par}

{\small{}Step 1: Initialize the inner iteration index $t=0$, $\left\{ p_{m,k,1}^{\left[0\right]},D_{j,k,1}^{\left[0\right]},p_{m,k,2}^{\left[0\right]},D_{j,k,2}^{\left[0\right]}\right\} $.}{\small\par}

{\small{}While $\left(\left|\mathcal{F}_{j}^{\left[t+1\right]}\right.\text{\textminus}\left.\mathcal{F}_{j}^{\left[t\right]}\right|\geq\delta_{I}\,\textrm{or}\,t\le T_{max}\right)$
do:\vspace{0.05cm}
}{\small\par}

{\small{}~~}%
\begin{tabular}{|>{\raggedright}p{8.5cm}}
{\small{}Step 2: Find the $\epsilon$-constraint solution $\left\{ p_{m,k,1}^{\left[t+1\right]},D_{j,k,1}^{\left[t+1\right]},p_{m,k,2}^{\left[t+1\right]},D_{j,k,2}^{\left[t+1\right]}\right\} $
of the following problem for given $\left\{ p_{m,k,1}^{\left[t\right]},D_{j,k,1}^{\left[t\right]},p_{m,k,2}^{\left[t\right]},D_{j,k,2}^{\left[t\right]}\right\} $,
and $r_{m,k,n}^{*}{}^{\left[m\right]}$ }{\small\par}

{\small{}~~~~~~~~$\left\{ p_{m,k,1}^{\left[t+1\right]},D_{j,k,1}^{\left[t+1\right]},p_{m,k,2}^{\left[t+1\right]},D_{j,k,2}^{\left[t+1\right]}\right\} =\textrm{Solving}\,\eqref{eq:30},\,\,$}{\small\par}

{\small{}Step 3: Update $\mathcal{F}_{j}^{\left[t+1\right]}$.}{\small\par}

{\small{}Step 4: }\textbf{\small{}Goto}{\small{} Step 2.}\tabularnewline
\end{tabular}{\small{}\vspace{0.05cm}
}{\small\par}

{\small{}end while.}{\small\par}

{\small{}Step 5: Return $\left\{ p_{m,k,1}^{*},D_{j,k,1}^{*},p_{m,k,2}^{*},D_{j,k,2}^{*}\right\} =\left\{ p_{m,k,1}^{\left[t+1\right]},D_{j,k,1}^{\left[t+1\right]},p_{m,k,2}^{\left[t+1\right]},D_{j,k,2}^{\left[t+1\right]}\right\} $
.}{\small\par}

{\small{}end}{\small\par}
\end{algorithm}
{\small{}
\begin{equation}
D_{j,k,n}\leq\log_{2}\left(1+\theta_{j,k,n}\right),\,\,\forall\,k,j,n,\label{eq:20}
\end{equation}
\begin{equation}
\theta_{j,k,n}\geq\frac{\mathrm{PL}\left(d_{m,k,j}\right)\left(2K_{m}-1\right)}{\sigma_{e}^{2}}p_{m,k,n}\varrho_{j,k,n},\label{eq:21}
\end{equation}
\begin{equation}
\varrho_{j,k,n}\geq W_{0}\left(\nu_{j,k,n}\right),\label{eq:22}
\end{equation}
\begin{equation}
\nu_{j,k,n}p_{m,k,n}^{2K_{m}}\geq\frac{\sigma_{e}^{2}}{\mathrm{PL}\left(d_{m,k,j}\right)^{2K_{m}}\left(2K_{m}-1\right)}\vartheta_{j,k,n},\label{eq:23}
\end{equation}
\begin{equation}
\vartheta_{j,k,n}\geq\left(\stackrel[\left(k',n'\right)\neq\left(k,n\right)]{}{\prod}\zeta_{j,k',n'}{\epsilon_{sop}}^{-1}\right)^{\frac{1}{2K_{m}-1}}.\label{eq:24}
\end{equation}
}{\small\par}

Now, since the non-convexity still persists within \eqref{eq:20}-\eqref{eq:24},
we attempt to approximate the non-convex factor at each iteration
by its first-order Taylor expansion at the $t^{st}$ SPCA iteration.
Following this approach, the affine approximation becomes straightforward
for each of \eqref{eq:20}-\eqref{eq:24}. Hence , we can replace
each non-convex constraint by its affine approximation, and thus the
equivalent convex form of \eqref{eq:19}-$C_{2}$ at the $t^{st}$
SPCA iteration may be formulated as:{\small{}
\begin{equation}
2^{D_{j,k,n}}\leq1+\theta_{j,k,n},\label{eq:25}
\end{equation}
\begin{equation}
\theta_{j,k,n}\geq\frac{\mathrm{PL}\left(d_{m,k,j}\right)\left(2K_{m}-1\right)}{\sigma_{e}^{2}}\varTheta^{[t]}\left(p_{m,k,n},\varrho_{j,k,n}\right),\label{eq:26}
\end{equation}
}{\footnotesize{}
\begin{equation}
\varrho_{j,k,n}\geq W_{0}\left(\nu_{j,k,n}^{[t]}\right)\left(\nu_{j,k,n}^{[t]}\left(1-W_{0}\left(\nu_{j,k,n}^{[t]}\right)\right)\right)^{-1}\left(\nu_{j,k,n}-\nu_{j,k,n}^{[t]}\right),\label{eq:27}
\end{equation}
}{\small{}
\begin{equation}
\Psi^{[t]}\left(\nu_{j,k,n},p_{m,k,n}^{2K_{m}}\right)\geq\frac{\sigma_{e}^{2}}{\mathrm{PL}\left(d_{m,k,j}\right)^{2K_{m}}\left(2K_{m}-1\right)}\vartheta_{j,k,n},\label{eq:28}
\end{equation}
}{\footnotesize{}
\begin{equation}
\log\left(\vartheta_{j,k,n}\right)\geq\frac{1}{\left(2K_{m}-1\right)}\left(\stackrel[\left(k',n'\right)\neq\left(k,n\right)]{}{\sum}\Lambda^{[t]}\left(\zeta_{j,k',n'}\right)-\log\left({\epsilon_{sop}}\right)\right),\label{eq:29}
\end{equation}
}$\forall\,k,j,n,$ respectively, {\small{}$\varTheta^{[t]}\left(p_{m,k,n},\varrho_{j,k,n}\right)\triangleq$
$\frac{1}{4}(p_{m,k,n}+\varrho_{j,k,n})^{2}$ $+\frac{1}{4}(p_{m,k,n}^{[t]}-\varrho_{j,k,n}^{[t]})^{2}$
$-\frac{1}{2}(p_{m,k,n}^{[t]}-\varrho_{j,k,n}^{[t]})$ $\left(p_{m,k,n}-\varrho_{j,k,n}\right),$
$\varGamma^{[m]}\left(D_{j,k,n}\right)\triangleq2^{D_{j,k,n}^{[t]}}$
$\left[1+\ln(2)\left(D_{j,k,n}-D_{j,k,n}^{[t]}\right)\right]$, $\Lambda^{[t]}\left(\zeta_{j,k',n'}\right)$
$\triangleq$ $\log\left(\zeta_{j,k',n'}^{[t]}\right)$ $+\frac{\zeta_{j,k',n'}-\zeta_{j,k',n'}^{[t]}}{\zeta_{j,k',n'}^{[t]}}$,}
and {\small{}$\Psi^{[t]}\left(\nu_{j,k,n},p_{m,k,n}^{2K_{m}}\right)$
$\triangleq$ $\nu_{j,k,n}^{[t]}$ $\left(p_{m,k,n}^{[t]}\right)^{2K_{m}}+$
$\left(p_{m,k,n}^{[t]}\right)^{\left(2K_{m}-1\right)}$ $\left[\left(p_{m,k,n}^{[t]}\right),\right.$
$\left(2K_{m}-1\right)$ $\left.\nu_{j,k,n}^{[t]}\right]$ $\times$
$\left[\nu_{j,k,n}-\nu_{j,k,n}^{[t]},\right.$ $\left.p_{m,k,n}-p_{m,k,n}^{[t]}\right]^{T}$.}{\small\par}

In order to arrive at \eqref{eq:27} from \eqref{eq:22}, we have
exploited the fact that since $W_{0}\left(x\right)$ is concave over
the interval of $\left(-e^{-1},\infty\right)$ and positive over $\left(1,\infty\right)$,
upon using the first order Taylor expansion of $W_{0}\left(x\right)$
we have {\small{}$W_{0}\left(\nu_{j,k,n}\right)\leq W_{0}$ $\left(\nu_{j,k,n}^{[t]}\right)$
$\left[\nu_{j,k,n}^{[t]}\right.$ $\left.\left(1-W_{0}\left(\nu_{j,k,n}^{[t]}\right)\right)\right]^{-1}$
$\left(\nu_{j,k,n}-\nu_{j,k,n}^{[t]}\right)$. }{\small\par}

\subsection{Overall Solution of the Original Problem \eqref{eq:10}}

Note that due to the decoding order constraint $\left\{ \Phi_{m}\right\} $,
it is challenging to find the optimal solution of problem \eqref{eq:10}.
To solve this problem, we first fix the decoding order $\Phi_{m}$
to obtain the optimal triplet $\left\{ r_{k,m,n}^{\ast}\geq0,\right.$
$D_{j,k,n}^{\ast}\geq0,$ $\left.p_{m,k,n}^{\ast}\geq0\right\} $
and then exhaustively search the entire set to find the optimal ${\Phi}_{m}^{\ast}$\footnote{Although it would also be beneficial to look for an optimal SIC-ordering
\cite{=00005B4=00005D,=00005B5=00005D}, as the UAV is assumed to
only have access to the AoA and distances, but not to the small-scale
fading parameters, we have not performed the SIC-ordering here and
left it for future works. It has been shown in {[}5{]} that for an
algorithm including $N$ initial points, the exhaustive SIC ordering
for a $K$ users uplink-RSMA imposes a tolerable computational complexity
of $\mathcal{O}\left(2^{K}+NK^{3}\left(2K!\right)/2^{K}\right)$.}. Upon assuming a fixed ${\Phi}_{m}$, we conceive a two-tier iterative
algorithm for attaining the overall $\epsilon$-constraint solution
$\text{\ensuremath{\left\{  p_{m,k,1}^{*},D_{j,k,1}^{*},p_{m,k,2}^{*},D_{j,k,2}^{*}\right\} } }$
in two different tiers. More explicitly, using the approximations
obtained in \eqref{eq:25}-\eqref{eq:29}, together with $\left\{ r_{k,m,1}^{\ast},r_{k,m,2}^{\ast}\right\} $
gleaned from the outer tier, the $\left(t+1\right)^{st}$ iteration
of the inner tier solves the following equivalent convex form of problem
\eqref{eq:19} for finding the $\epsilon$-constraint solution as
\eqref{eq:30},
\begin{figure*}[tbh]
\begin{equation}
\underset{1\text{\ensuremath{\le}}j\text{\ensuremath{\le}}J}{\min}\left(\underset{\mathbf{X}}{\max}\left\{ \mathcal{F}_{j}\right\} \right),\,\,\label{eq:30}
\end{equation}
s.t.

\begin{tabular}{>{\raggedright}p{8cm}}
$C_{1}:$ $\stackrel[\left(k',n'\right)\in\Phi_{m,k,n}]{}{\sum}\log_{2}\left(p_{m,k',n'}\right)\leq\psi_{k},\,\forall\,k,n$,\vspace{0.1cm}
\tabularnewline
\end{tabular}%
\begin{tabular}{>{\raggedright}p{7cm}}
$C_{2-1}:$ $\varGamma^{[t]}\left(D_{j,k,n}\right)\geq1+\theta_{j,k,n},\,\,\forall\,k,j,n$,\vspace{0.1cm}
\tabularnewline
\end{tabular}

\begin{tabular}{>{\raggedright}p{16cm}}
$C_{2-2}:$ $\theta_{j,k,n}\geq\frac{\mathrm{PL}\left(d_{m,k,j}\right)\left(2K_{m}-1\right)}{\sigma_{e}^{2}}\varTheta^{[t]}\left(p_{m,k,n},\varrho_{j,k,n}\right),\,\,\forall\,k,j,n$,

\vspace{0.1cm}
$C_{2-3}:$ $\varrho_{j,k,n}\geq W_{0}\left(\nu_{j,k,n}^{[t]}\right)\left(\nu_{j,k,n}^{[t]}\left(1-W_{0}\left(\nu_{j,k,n}^{[t]}\right)\right)\right)^{-1}\left(\nu_{j,k,n}-\nu_{j,k,n}^{[t]}\right),\,\,\forall\,k,j,n$,

\vspace{0.1cm}
$C_{2-4}:$ $\Psi^{[t]}\left(\nu_{j,k,n},p_{m,k,n}^{2K_{m}}\right)\geq\frac{\sigma_{e}^{2}}{\mathrm{PL}\left(d_{m,k,j}\right)^{2K_{m}}\left(2K_{m}-1\right)}\vartheta_{j,k,n},\,\,\forall\,k,j,n$,

\vspace{0.1cm}

$C_{2-5}:$ $\log\left(\vartheta_{j,k,n}\right)\geq\frac{1}{\left(2K_{m}-1\right)}\left(\stackrel[\left(k',n'\right)\neq\left(k,n\right)]{}{\sum}\Lambda^{[t]}\left(\zeta_{j,k',n'}\right)-\log\left({\epsilon_{sop}}\right)\right),\,\,\forall\,k,j,n$,

\vspace{0.1cm}
\tabularnewline
\end{tabular}

\begin{tabular}{>{\raggedright}p{17.5cm}}
$C_{3}:$ $r_{m,k,n}^{*}\geq D_{j,k,n},\,\,\forall\,k,n$,~~~~~$C_{4}:$
$\stackrel[n=1]{2}{\sum}p_{m,k,n}\leq P_{m,k},\,\,\forall\,k$.

\vspace{0.1cm}
\tabularnewline
\hline 
\end{tabular}
\end{figure*}
{\small{} }where $\mathbf{X}\triangleq\left(\boldsymbol{x},\left\{ p_{m,k,1},D_{j,k,1},\ p_{m,k,2},D_{j,k,2}\right\} _{k=1}^{K_{M}}\right)$,
and $\boldsymbol{x}\triangleq\left\{ \theta_{j,k,n},\varrho_{j,k,n},\nu_{j,k,n},\vartheta_{j,k,n},\nu_{j,k,n}\right\} $.
Upon the $\epsilon$-constraint point $\left\{ p_{m,k,1}^{*},D_{j,k,1}^{*},p_{m,k,2}^{*},D_{j,k,2}^{*}\right\} $
found by the inner loop, the $\left(q+1\right)^{st}$ iteration of
the outer loop finds $\epsilon$-constraint solution $\left\{ r_{m,k,1}^{*}{}^{\left[q+1\right]},r_{m,k,2}^{*}{}^{\left[q+1\right]}\right\} $,
given by \eqref{eq:31}, 
\begin{figure*}[tbh]
{\small{}
\begin{gather}
{\left\{ p_{m,k,1}^{*},D_{j,k,1}^{*},p_{m,k,2}^{*},D_{j,k,2}^{*}\right\} }\,\textrm{obtained\,from}\,\eqref{eq:30}\nonumber \\
\Downarrow\,\,\,\Uparrow\nonumber \\
\textrm{update\,}\,r_{m,k,n}^{*}{}^{\left[q+1\right]}=\log_{2}\left(1+\frac{2A}{\sigma_{m}^{2}}W_{0}\left(\left[\xi{\epsilon_{cop}}^{-1}\stackrel[\left(k',n'\right)\in\Phi_{m,k,n}]{}{\prod}\left(2\lambda_{m,k',n'}^{*^{-1}}\right)\right]^{\frac{1}{A}}\right)\right),\label{eq:31}
\end{gather}
\_\_\_\_\_\_\_\_\_\_\_\_\_\_\_\_\_\_\_\_\_\_\_\_\_\_\_\_\_\_\_\_\_\_\_\_\_\_\_\_\_\_\_\_\_\_\_\_\_\_\_\_\_\_\_\_\_\_\_\_\_\_\_\_\_\_\_\_\_\_\_\_\_\_\_\_\_\_\_\_\_\_\_\_\_\_\_\_\_\_\_\_\_\_\_\_\_\_\_\_\_\_}
\end{figure*}
where the superscript \textquotedblleft $*$\textquotedblright{} represents
the final iteration of the inner loop. Since the proposed method consists
of two layers of iterations, the stopping criterion of each layer
depends on the relative change of the two consecutive of values. Therefore,
the outer loop proceeds to the next iteration and runs until $\left|r_{m,k,n}^{*}{}^{\left[q+1\right]}-r_{m,k,n}^{*}{}^{\left[q\right]}\right|\le\epsilon$
is met or the maximum affordable number of iterations $Q_{max}$ is
reached. To find the $\epsilon$-constraint solution $\text{\ensuremath{\left\{  p_{m,k,1}^{*},p_{m,k,2}^{*},D_{j,k,n}^{*}\right\} } }$,
the problem (\eqref{eq:30}) is solved using the classic SPCA in another
iterative process of the inner loop. In particular, the inner iterations
are continued until the stopping criterion of $\left|\mathcal{F}_{j}^{\left[t+1\right]}\right.$
$\text{\textminus}$ $\left.\mathcal{F}_{j}^{\left[t\right]}\right|\le\delta_{I}$\footnote{\_\_\_\_\_\_\_\_\_\_\_\_\_\_\_\_\_\_\_\_\_\_\_\_\_\_\_\_\_\_\_\_\_\_\_\_\_\_\_\_\_\_\_\_\_\_\_\_\_\_\_\_\_\_\_\_\_\_\_

Note that $\mathcal{F}_{j}^{\left[t\right]}$ represents the value
of $\mathcal{F}_{j}$ at iteration $t^{th}$ .} is satisfied at the $\left(t+1\right)^{st}$ iteration or the maximum
affordable number of iterations $T_{max}$ is reached. The proposed
two-tier scheme is presented in Algorithm I.

\subsection{Complexity Analysis}

In Algorithm 1, the major complexity lies in solving problem \eqref{eq:30}.
According to Algorithm 1, a globally near-optimal solution of problem
\eqref{eq:30} is obtained via solving a series of convex problems
with different initial points and decoding order strategies. Considering
that the dimension of the variables in problem \eqref{eq:30} is $\mathcal{L}_{m}=\text{5\ensuremath{\left(1+J\right)}}K_{m}$,
the worst-case complexity order of solving the convex problem in Step
2 of inner-loop by using the standard interior point method is given
by $\mathcal{O}\left(\left(\frac{\mathcal{L}_{m}-1}{2}\right)^{3}\right)$
\cite[Pages 487, 569]{=00005B30=00005D}. Since each cluster consists
of $K_{m}$ users and each user transmits a superposition of two messages
(there are $2K_{m}$ messages for the $m^{th}$ cluster), the decoding
order set $\mathbf{\Pi}_{m}$ consists of $\frac{\left(2K_{m}\right)!}{2^{K_{m}}}$
elements. Therefore, the total complexity of solving problem \eqref{eq:30}
at each iteration is given by $\mathcal{O}\left(\left(\frac{\mathcal{L}_{m}-1}{2}\right)^{3}\frac{\left(2K_{m}\right)!}{2^{K_{m}}}\right)$.
In practice, we consider small $K_{m}$ to reduce the SIC complexity,
so that the computational complexity of Algorithm 1 remains practical.
To deal with a large number of users, we can increase the number of
clusters and the users can be classified into different clusters,
each having a small number of users. 

\subsection{Convergence Analysis}

In this section we establish a convergence analysis for the SPCA algorithm.
Since the original problem \eqref{eq:15} is non-convex, it is not
possible to prove convergence to a global minimum but rather convergence
to $\mathrm{KKT}$ points under some regularity conditions. We use
the following simple and technical lemmas will be used in the convergence
proof. For simplicity we define $\varOmega\triangleq\textrm{feasible\,set\,of }$
\eqref{eq:15}, $\varOmega^{\left[t\right]}\triangleq\textrm{feasible\,set\,of }$
\eqref{eq:30}\footnote{\_\_\_\_\_\_\_\_\_\_\_\_\_\_\_\_\_\_\_\_\_\_\_\_\_\_\_\_\_\_\_\_\_\_\_\_\_\_\_\_\_\_\_\_\_\_\_\_\_\_\_\_\_\_\_\_\_\_\_\_

In this paper, we consider the standard form of generic optimization
problem as follow \cite{36}:
\[
\begin{array}{cc}
\mathit{\textrm{min}\:\:}f(\mathbf{x})\\
\textrm{s.t.}\:\mathit{c_{j}(\mathbf{x)\leq0}}, & \forall\,j=1,2,...,m\\
\mathbf{x}\in\mathbb{R}^{n}
\end{array}
\]
where $\mathit{f(\mathbf{x})\:\textrm{and}\:c_{j}(\mathbf{x}),\:\forall\,j=1,...,m}$
are all continuously differentiable objective and constraint functions
over $\mathit{\mathbb{R}^{n}}$, respectively. Also, we assume that
the function $\mathit{f(\mathbf{x})}$ and the last $\mathit{m-p}$
constraint functions $\mathit{c_{p+1}(\mathbf{x}),...,c_{m}(\mathbf{x})}$
($\mathit{p\le m}$) are convex over $\mathit{\mathit{\mathbb{R}^{n}}}$.
Therefore, the \textquotedblleft \textit{non-convex part}\textquotedblright{}
of the problem is due to the nonconvexity of the first $\mathit{p}$
constraint functions $\mathit{c_{1}(\mathbf{x}),...,c_{p}(\mathbf{x})}$.
The case $\mathit{p=m}$ corresponds to the case when all the constraints
are non-convex. In addition, suppose that for every $j=1,...,p$,
$c_{j}(\mathbf{x})$ has a continuous convex upper estimate function
$C_{j}:\mathbb{R}^{n}\times\mathbb{Y}\rightarrow\mathbb{R}$, specifically,
assume that there exists a set $\mathbb{Y}\subseteq\mathbb{R}^{r}$
($r$ is a positive integer), such that $c_{j}(\mathbf{x})\leq C_{j}(\mathbf{x},\boldsymbol{\rho}),$
$\forall\,$$\mathbf{x}$ $\in\mathbb{R}^{n},\:\forall\,\boldsymbol{\rho}\in\mathbb{Y},$
where for a fixed $\boldsymbol{\rho}$ the function $C_{j}(\mathbf{.},\boldsymbol{\rho})$
is convex and continuously differentiable. The basic idea of SPCA
is that at each iteration $i$, we replace each non-convex functions
$c_{j}(\mathbf{x}),\:\forall\,j=1,...,p$ by the upper convex approximation
function $C_{j}(\mathbf{x},\boldsymbol{\rho})$ for some appropriately
chosen parameter vector $\boldsymbol{\rho}$. Thus, at step $i\:(i\ge1)$
we need to solve the following equivalent convex problem:
\[
\qquad\:\quad\begin{array}{cc}
\mathit{\textrm{min}\:\:}f(\mathbf{x})\\
\textrm{s.t.}\:\mathit{C_{j}(\mathbf{x},\boldsymbol{\rho})\leq0}, & \forall\,j=1,2,...,p\\
\:\mathit{c_{i}(\mathbf{x)\leq0}}, & \quad\quad\:\:\forall\,j=p+1,p+2,...,m\\
\mathbf{x}\in\mathbb{R}^{n}
\end{array}
\]
} for $t^{th}$ iteration.
\begin{lem}
Let $\mathcal{D}$ : $\mathbb{R}^{n}\rightarrow\mathbb{R}$ be a strictly
convex and differentiable function on a nonempty convex set $S\subseteq\mathbb{R}^{n}$.
Then $\mathcal{D}$ is strongly convex on the set $S$.
\end{lem}
\begin{IEEEproof}
See \cite{34}.
\end{IEEEproof}
\begin{lem}
Suppose $\left\{ {\mathbf{X}^{\left[t\right]}}\right\} $ be the sequence
generated by the SPCA method. Then for every $t\ge0$: \textbf{i).}
\textup{$\varOmega^{\left[t\right]}$} $\subseteq\varOmega,$~ \textbf{ii).}
\textup{${\mathbf{x}^{\left[t\right]}}$} $\in$ \textup{$\varOmega^{\left[t\right]}$}
$\cap$ \textup{$\varOmega^{\left[t+1\right]}$},~\textbf{ iii).}
\textup{${\mathbf{X}^{\left[t\right]}}$} is a feasible point of \eqref{eq:15},
\textbf{~iv).} \textup{$\mathcal{F}_{j}^{\left[t+1\right]}\,\le\,\mathcal{F}_{j}^{\left[t\right]}$}.
\end{lem}
\begin{IEEEproof}
See \cite{34}.
\end{IEEEproof}
\begin{lem}
The sequence $\left\{ \mathcal{F}_{j}^{\left[t\right]}\right\} $
converges.
\end{lem}
\begin{IEEEproof}
See \cite{34}.
\end{IEEEproof}
Recall that a feasible solution $\mathbf{X}^{\logof}$ of a optimization
problem is $\mathit{regular}$ if the set of gradients of the active
constraints at $\mathbf{X}^{\logof}$ is linearly independent \cite{35}.
If ${\mathbf{X}^{\left[t\right]}}$ converges to a regular point $\mathbf{X}^{\logof}$,
then $\mathbf{X}^{\logof}$ is a $\mathrm{KKT}$ point of problem
\eqref{eq:15}. By $\textrm{Lemma}$$\textrm{2}$ it follows that
the strictly convex objective function $\mathcal{F}_{j}$ is also
strongly convex on the convex feasible set $\varOmega^{\left[t+1\right]}$.
In particular, there exists $\vartheta>0$ such that for all $k\ge0$
we have:{\small{}
\begin{gather}
\mathcal{F}_{j}^{\left[t\right]}-\mathcal{F}_{j}^{\left[t+1\right]}\ge\left(\mathbf{X}^{\left[t\right]}-\mathbf{X}^{\left[t+1\right]}\right)^{\text{T}}\nabla\mathcal{F}_{j}^{\left[t+1\right]}+\vartheta\left\Vert \mathbf{X}^{\left[t\right]}-\mathbf{X}^{\left[t+1\right]}\right\Vert ^{2},\label{eq:32}
\end{gather}
}since ${\mathbf{X}^{\left[t\right]}}$ is a feasible point of \eqref{eq:30}
(by Lemma 3), and ${\mathbf{X}^{\left[t+1\right]}}$ is its optimum,
then from the optimality conditions for $\left(t+1\right)^{th}$ iteration
of \eqref{eq:30} (see \cite[proposition 2.1.2]{36}), we obtain $\left(\mathbf{X}^{\left[t\right]}-\mathbf{X}^{\left[t+1\right]}\right)^{\text{T}}\nabla\mathcal{F}_{j}^{\left[t+1\right]}\geq0$,
which combined with \eqref{eq:32} yields: 
\begin{equation}
\mathcal{F}_{j}^{\left[t\right]}-\mathcal{F}_{j}^{\left[t+1\right]}\ge\vartheta\left\Vert \mathbf{X}^{\left[t\right]}-\mathbf{X}^{\left[t+1\right]}\right\Vert ^{2}\label{eq:33}
\end{equation}
 By $\textrm{Lemma}$ 3, the sequence $\left\{ \mathcal{F}_{j}^{\left[t\right]}\right\} $
converges and thus the inequality \eqref{eq:33} implies that $\left\Vert \mathbf{X}^{\left[t\right]}-\mathbf{X}^{\left[t+1\right]}\right\Vert \rightarrow0$.
Let be $\mathbf{X}^{\diamondsuit}\triangleq\left({\color{red}\boldsymbol{x}}^{\diamondsuit},\left\{ p_{m,k,1}^{\diamondsuit},D_{j,k,1}^{\diamondsuit},p_{m,k,2}^{\diamondsuit},D_{j,k,2}^{\diamondsuit}\right\} _{k=1}^{K_{M}}\right)$
an accumulation point of the sequence $\left\{ \mathbf{X}^{\left[t\right]}\right\} $,
we will show that $\mathbf{X}^{\diamondsuit}$ is a $\mathrm{KKT}$
point. Since $\mathbf{X}^{\diamondsuit}$ is an accumulation point
of $\left\{ \mathbf{X}^{\left[t\right]}\right\} $, there exists a
subsequence $\left\{ \mathbf{X}^{\left[t_{n}\right]}\right\} $ such
that $\mathbf{X}^{\left[t_{n}\right]}\rightarrow\mathbf{X}^{\diamondsuit}$
when the iteration number $n\rightarrow\infty$. Regarding the limit
point $\mathbf{X}^{\diamondsuit}$ , we can make the following statement.
\begin{cor}
The accumulation point \textup{$\mathbf{X}^{\diamondsuit}$} of the
sequence $\left\{ \mathbf{X}^{\left[t\right]}\right\} $ generated
by the proposed SPCA method is a KKT point of the \eqref{eq:30}.
\end{cor}
\begin{IEEEproof}
We know from \cite{35} that there exist Lagrangian multipliers $\lambda_{i}^{*}$
together with the accumulation point $\mathbf{X}^{\diamondsuit}$
that satisfy the following $\mathrm{KKT}$\textquoteright s necessary
and sufficient condition for optimality of convex problem \cite[Sec 5.5]{35},
at \eqref{eq:34} $\left\{ \gamma_{i}\right\} _{i=1}^{8}$ denote
the lagrangian multipliers of problem \eqref{eq:30}. If we choose
$\gamma_{i}=\lambda_{i}$ for $i=1,....,8$ we conclude that the point
$\mathbf{X}^{\diamondsuit}$ also satisfy so, we proved that if the
sequence $\left\{ \mathbf{X}^{\left[t\right]}\right\} $ generated
by the SPCA method converges to a regular point $\mathbf{X}^{\diamondsuit}$,
then $\mathbf{X}^{\diamondsuit}$ is a $\mathrm{KKT}$ point of the
SPCA problem \eqref{eq:30}. It has already been shown that the point
$\mathbf{X}^{\diamondsuit}$ is a $\mathrm{KKT}$ point (stationary
point) of the SPCA problem \eqref{eq:30}. This stationary point cannot
be saddle point, since the objective function $\mathcal{F}_{j}$ is
strictly convex function and twice-continuously differentiable in
the variable $\mathbf{X}$. By a simple contradiction method, we can
also show that the point $\mathbf{X}$ cannot be a local maximum \cite{35}.
\begin{figure*}[tbh]
{\small{}
\[
\nabla\mathcal{F}_{j}+\gamma_{1}\nabla\left(\stackrel[\left(k',n'\right)\in\Phi_{m,k,n}]{}{\sum}\log_{2}\left(p_{m,k',n'}\right)-\psi_{k}\right)+\gamma_{2}\nabla\left(1+\theta_{j,k,n}-\varGamma^{[t]}\left(D_{j,k,n}\right)\right)+\gamma_{3}\nabla\Biggl(\frac{\mathrm{PL}\left(d_{m,k,j}\right)\left(2K_{m}-1\right)}{\sigma_{e}^{2}}
\]
\[
+\varTheta^{[t]}\left(p_{m,k,n},\varrho_{j,k,n}\right)-\theta_{j,k,n}\Biggr)+\gamma_{4}\nabla\left(W_{0}\left(\nu_{j,k,n}^{[t]}\right)\left(\nu_{j,k,n}^{[t]}\left(1-W_{0}\left(\nu_{j,k,n}^{[t]}\right)\right)\right)^{-1}\left(\nu_{j,k,n}-\nu_{j,k,n}^{[t]}\right)-\varrho_{j,k,n}\right)+
\]
\[
\gamma_{5}\nabla\left(\frac{\sigma_{e}^{2}}{\mathrm{PL}\left(d_{m,k,j}\right)^{2K_{m}}\left(2K_{m}-1\right)}\vartheta_{j,k,n}-\Psi^{[t]}\left(\nu_{j,k,n},p_{m,k,n}^{2K_{m}}\right)\right)+\gamma_{6}\nabla\left(D_{j,k,n}-r_{m,k,n}^{*}\right)+
\]
\begin{equation}
\gamma_{7}\nabla\Biggl(\frac{1}{\left(2K_{m}-1\right)}\left(\stackrel[\left(k',n'\right)\neq\left(k,n\right)]{}{\sum}\Lambda^{[t]}\left(\zeta_{j,k',n'}\right)-\log\left({\epsilon_{sop}}\right)\right)-\log\left(\vartheta_{j,k,n}\right)\Biggr)+\gamma_{8}\nabla\left(\stackrel[n=1]{2}{\sum}p_{m,k,n}-P_{m,k}\right)=0.\label{eq:34}
\end{equation}
}{\small\par}

{\small{}\_\_\_\_\_\_\_\_\_\_\_\_\_\_\_\_\_\_\_\_\_\_\_\_\_\_\_\_\_\_\_\_\_\_\_\_\_\_\_\_\_\_\_\_\_\_\_\_\_\_\_\_\_\_\_\_\_\_\_\_\_\_\_\_\_\_\_\_\_\_\_\_\_\_\_\_\_\_\_\_\_\_\_\_\_\_\_\_\_\_\_\_\_\_\_\_\_\_\_\_\_\_}{\small\par}
\end{figure*}
\end{IEEEproof}

\section{Simulation Results}

In this section, we evaluate the secure transmission performance of
the proposed algorithm through simulations. Each point in the figures
is obtained by averaging over $150$ simulation trials. Unless otherwise
specified, the simulation setup is as follows throughout this section.
In the scenario investigated a UAV hovers above the users to provide
communication services. Explicitly, the UAV has a coverage radius
of $R_{UAV}=800\,\textrm{m}$ and altitude of $H_{i}=140\,\textrm{m}$.
The path-loss model in the UAV network includes both LoS and non-LoS
links associated with the path-loss exponents of $\mathcal{L}_{m,k}=2\,$
and $\mathcal{N}_{m,k}=3.5$, respectively. There are $M=3$ clusters,
$100$ users associated with $P_{i,k}=\frac{P}{\stackrel[m=1]{M}{\sum}K_{m}}\,$,
where $P$ is the total power budget, and $J=3$ $Eve$s randomly
distributed in the whole system between $1\,\textrm{m}$ and $800\,\textrm{m}$.
Once the large-scale fading parameters are generated, they are assumed
to be known and fixed throughout the simulations. The small-scale
fading vectors of all users and $Eve$s are independently generated
according to $\mathcal{CN}(0,\,\mathbf{I}_{N_{t}})$.  The noise power
at each user and eavesdropper is set to $\sigma_{m}^{2}=\sigma_{e}^{2}=0\,\textrm{dB}$.
Moreover, we set $T_{max}=Q_{max}=20$, $N_{t}=5$, $\epsilon_{cop}=\epsilon_{sop}=0.1$,
and the maximum threshold value used for the termination of Algorithm
1 is set to $\delta_{I}=10^{-2}$. The maximum tolerance of $\epsilon=10^{-3}$
is assumed for the termination criterion used in Algorithm 1. 
\begin{figure*}[tbh]
\raggedright{}%
\begin{minipage}[c][1\totalheight][t]{0.6\columnwidth}%
\begin{center}
\includegraphics[scale=0.62]{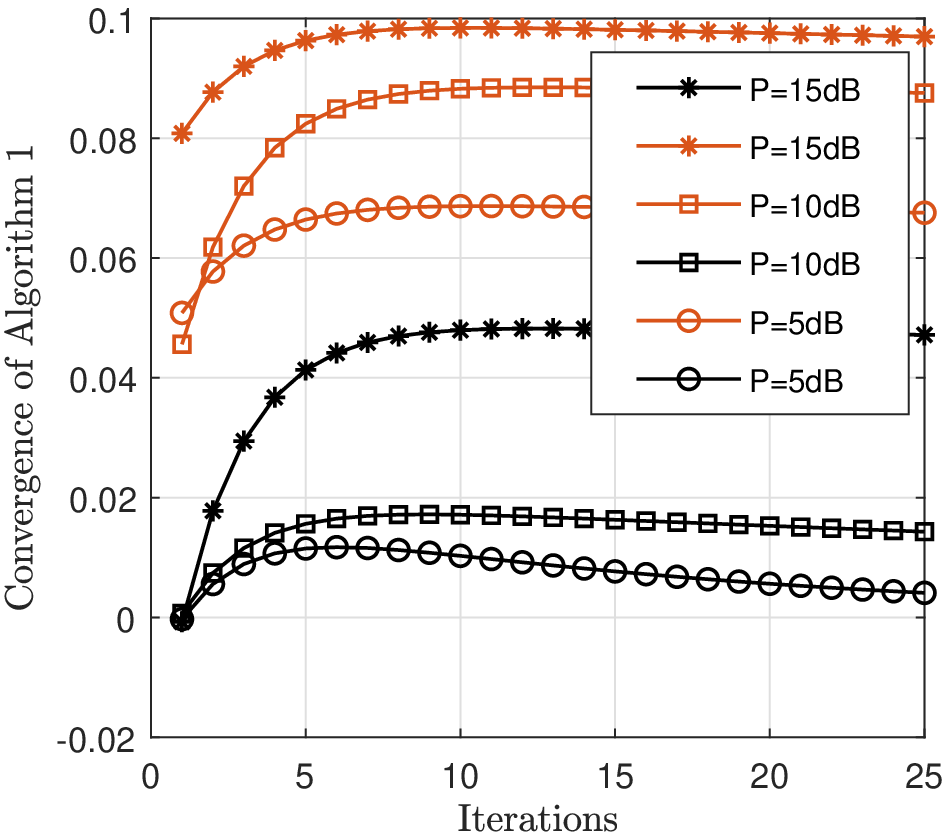}\caption{\label{fig:The-Convergence-of}The Convergence of Algorithm 1, black-line
refer to $\mathcal{F}_{j}^{\left[t\right]}$ and red-line refer to
$r_{m,k,n}^{*}{}^{\left[q\right]}$ for different values of $P$.}
\par\end{center}%
\end{minipage}~~~~~%
\begin{minipage}[c][1\totalheight][t]{0.6\columnwidth}%
\begin{center}
\includegraphics[scale=0.71]{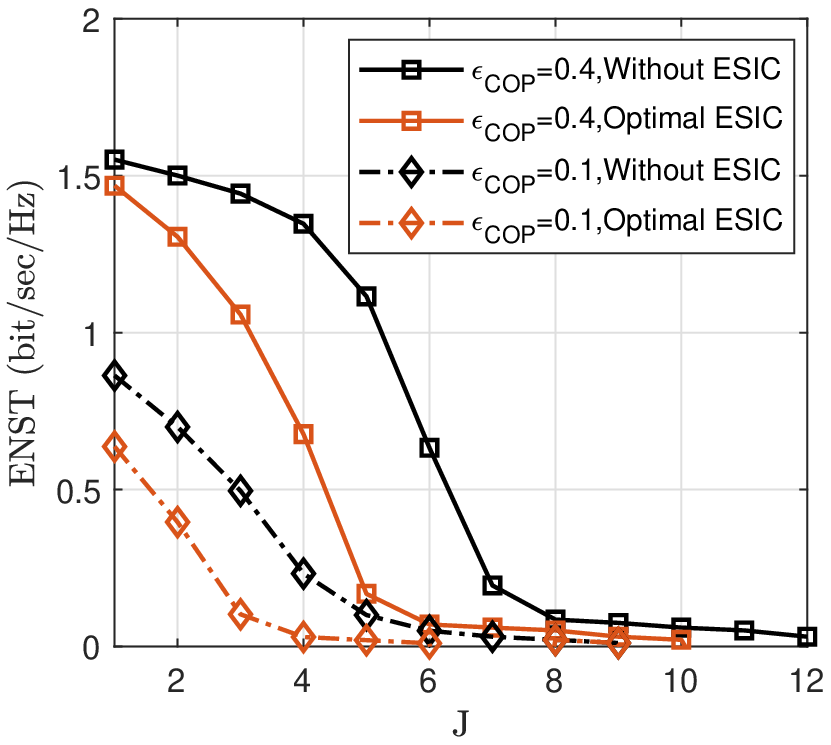}\caption{\label{fig:ENST-versus--3}ENST versus $J$ for different values of
$\epsilon_{cop}$ and with/without optimal $Eve$ SIC ordering.}
\par\end{center}%
\end{minipage}~~~~%
\begin{minipage}[c][1\totalheight][t]{0.65\columnwidth}%
\begin{center}
\includegraphics[viewport=0bp 0bp 305bp 260bp,clip,scale=0.58]{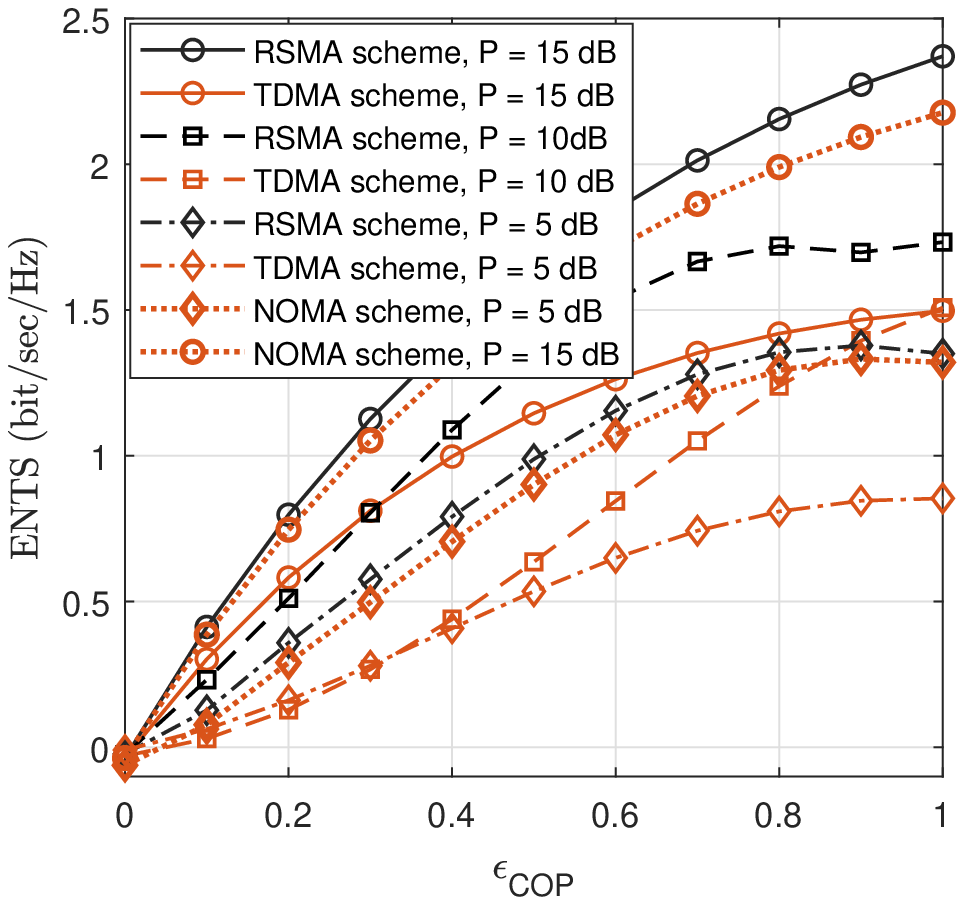}\caption{\label{fig:ENST-versus-}ENST versus $\epsilon_{cop}$ for different
values of $P$.}
\par\end{center}%
\end{minipage}
\end{figure*}

Firstly, in Fig. \ref{fig:The-Convergence-of} we demonstrate the
convergence of Algorithm $1$ is solving \eqref{eq:18} and \eqref{eq:19}
for different values of $P$. The convergence of the inner loop of
Algorithm 1 in terms of updating $\mathcal{F}_{j}^{\left[t\right]}$
is shown by black-lines. It is observed that the inner loop converges
within $6$ iterations for different values of $P$, which corroborates
the convergence of \eqref{eq:30}. When fixing the number of users,
the power $P_{i,k}$ allocated to each user increases upon increasing
$P$. As a result, the OF value of $\mathcal{F}_{j}^{\left[t\right]}$
in \eqref{eq:30} increases as $P$ increases. On the other hand,
the convergence of the outer loop of Algorithm $1$ is characterized
by the red-lines. Observe that the algorithm used for solving \eqref{eq:18}
converges after a maximum of $10$ iterations under different values
of $P$. Fig. \ref{fig:ENST-versus--3} shows the ENST of the proposed
scheme versus $J$ for different values of $\epsilon_{cop}$. Naturally,
upon increasing $J$, the performance degrades due to having more
$Eves$ in the system, but using a higher $\epsilon_{cop}$ would
increase ENST and compensate for the performance loss. Additionally,
as another important observation, without performing SIC at \textit{Eve}
(ESIC), the \textit{Eve}'s rate is increased, resulting in ENST enhancement. 

To show the performance advantages of the proposed scheme by employing
the RSMA scheme, we compare it to both power-domain (PD) NOMA and
TDMA. In PD-NOMA, the UAV first decodes the messages of users having
high channel gains and then decodes the messages of users with low
channel gains by subtracting the interference imposed by the previously
decoded high-gain user. In TDMA, each user will be assigned a fraction
of time to use the whole bandwidth. Let $\alpha_{m,k}=\frac{1}{K_{m}}$
be the fraction of time allocated to $U_{m,k}$. Then the data rate
of $U_{m,k}$ becomes $C_{m,k}^{TDMA}=\alpha_{m,k}\log_{2}\left(1+\frac{p_{m,k,n}\left|\mathbf{w}_{m}^{H}\mathbf{h}_{m,k}\right|^{2}}{\sigma_{m}^{2}}\right)$.
Observe from Fig. \ref{fig:ENST-versus-} that RSMA always achieves
a better performance than PD-NOMA and TDMA. Moreover, the ENST gain
of the proposed scheme over TDMA becomes more prominent as $P$ and
$\epsilon_{cop}$ increases.
\begin{figure*}[tbh]
\raggedright{}%
\begin{minipage}[c][1\totalheight][t]{0.65\columnwidth}%
\begin{center}
\includegraphics[viewport=0bp 0bp 248bp 246bp,clip,scale=0.7]{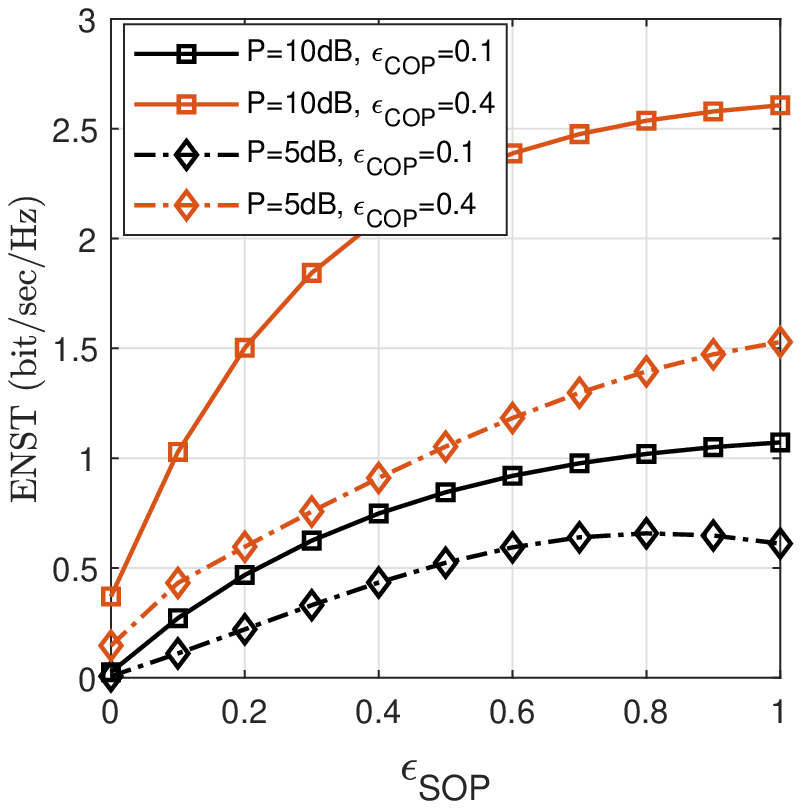}\caption{\label{fig:ENST-versus--1}ENST versus $\epsilon_{sop}$ for different
values of $P$ and $\epsilon_{cop}$.}
\par\end{center}%
\end{minipage}~%
\begin{minipage}[c][1\totalheight][t]{0.65\columnwidth}%
\begin{center}
~\includegraphics[viewport=0bp 0bp 365bp 317bp,clip,scale=0.5]{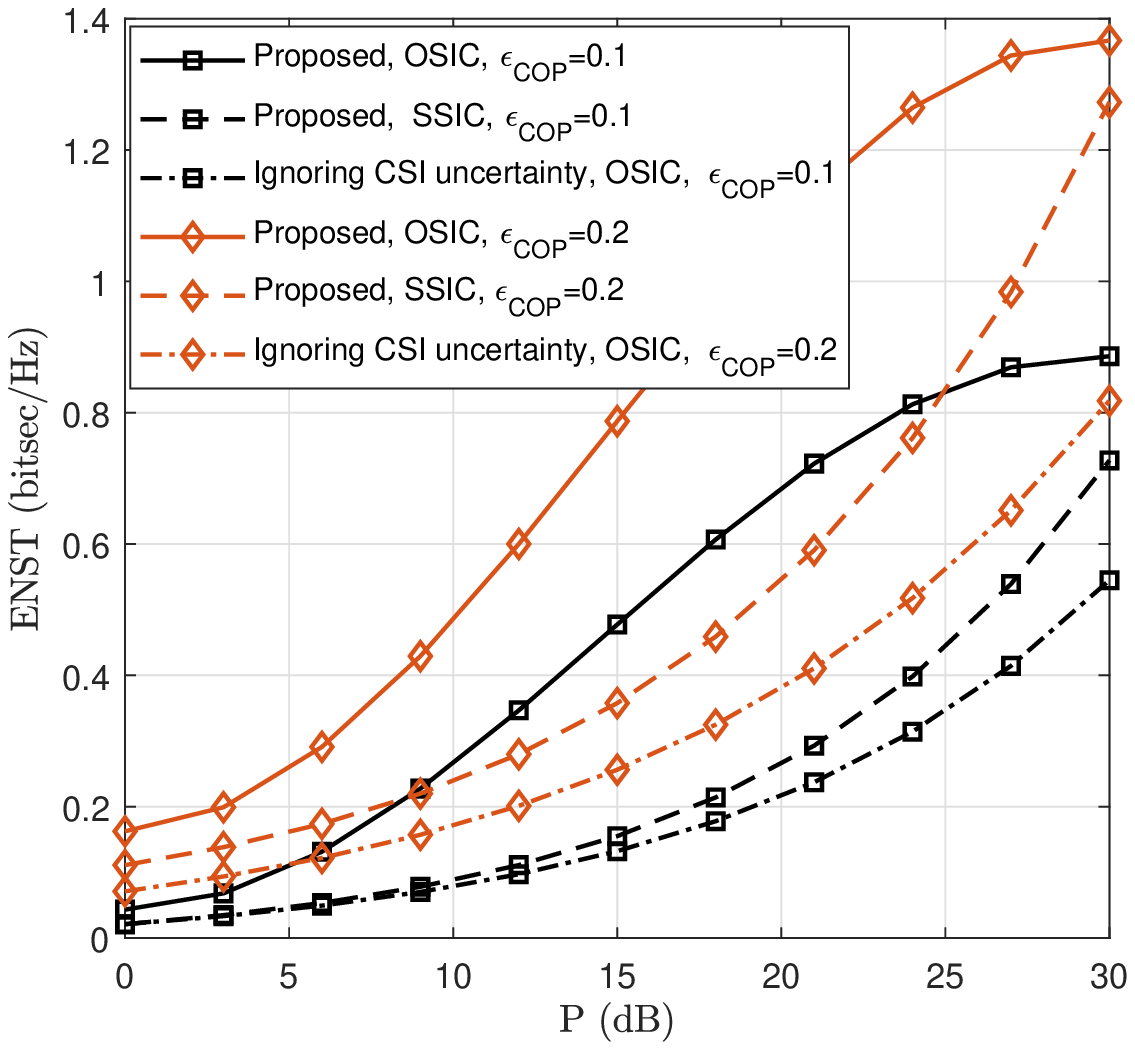}\caption{\label{fig:ENST-versus--2}ENST versus $P$ for different values of
$\epsilon_{cop}$ and SIC method.}
\par\end{center}%
\end{minipage}~~~~%
\begin{minipage}[c][1\totalheight][t]{0.65\columnwidth}%
\begin{center}
\includegraphics[viewport=0bp 0bp 305bp 260bp,clip,scale=0.63]{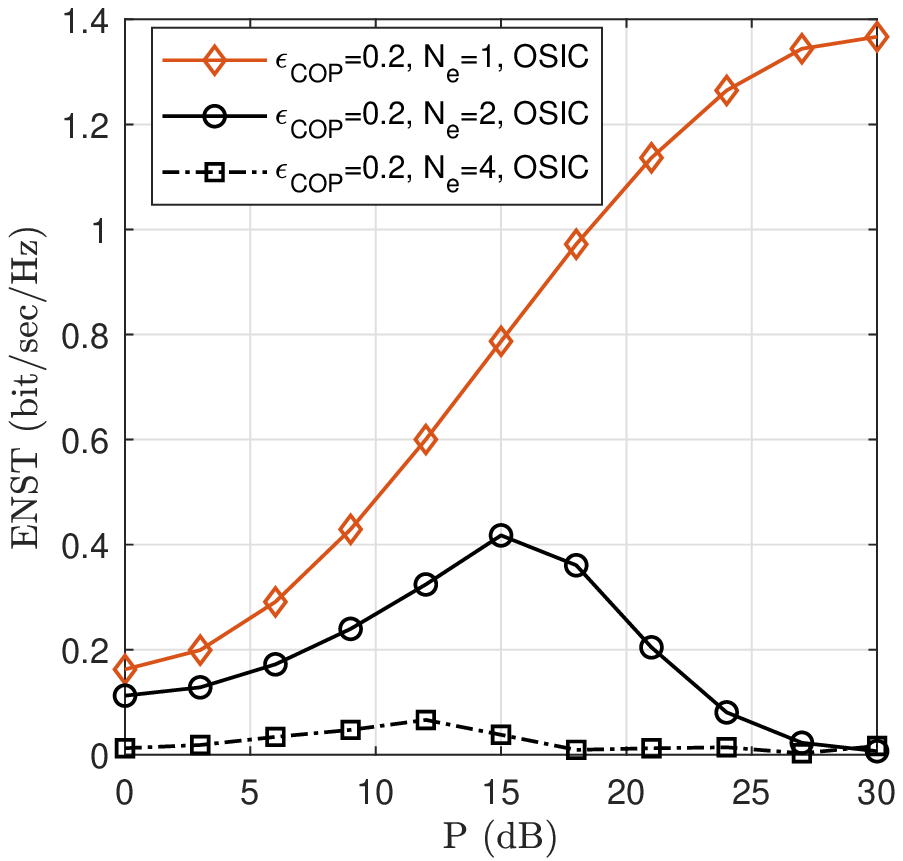}\caption{\label{fig:ENST-versus--4}ENST versus $\epsilon_{cop}$ for different
values of $N_{e}$.}
\par\end{center}%
\end{minipage}
\end{figure*}

The corresponding ENST versus $\epsilon$ plot is provided in Fig.
\ref{fig:ENST-versus--1}, where the proposed scheme achieves a significantly
higher ENST upon increasing $P$, $\epsilon_{sop}$ and $\epsilon_{cop}$.
A heuristic explanation of this phenomenon is that increasing both
the connection and secrecy outage threshold tends to relax the constraints
of \eqref{eq:30}, and decrease the lower bound of \eqref{eq:19}-$C_{2}$,
which in turn increases the ENST. 

Finally, to show the importance of considering SIC ordering as well
as imperfect CSIT, we compare the proposed scheme for both optimal
SIC (OSIC) and Sub-optimal SIC (SSIC) ordering. Additionally, we also
consider RSMA with perfect CSIT. To make a fair comparison, we simulate
all schemes under the same security requirement. The ENST versus $P$
trends recorded for different values of $\epsilon_{cop}$ are provided
in Fig. \ref{fig:ENST-versus--2}, where the proposed scheme always
achieves significantly higher ENST than RSMA ignoring CSIT uncertainty.
Fig. \ref{fig:ENST-versus--2} suggests that using bigger $\epsilon_{cop}$
would increase the ENST and mitigate the performance loss of SSIC.

Finally, Fig. 9 illustrates the ENST versus $N_{e}$. This figure
indicates that increasing the number of the receive antennas at $Eve$,
the system's secrecy performance is degraded due to $Eve$'s improved
ability to eavesdrop and infer from common message. Interestingly,
our proposed scheme still shows considerable robustness against a
multiple antenna-aided $Eve$, hence we can achieve non-zero ENST.

\section{Conclusions}

In this article, we proposed secure RSMA uplink transmission under
imperfect CSIT for a UAV-BS network, in which RSMA is employed by
each legitimate users for secure transmission under large-scale uplink
access. To characterize the performance of this system, an efficient
block coordinate decent algorithm was proposed for maximizing the
effective network secrecy throughput under the constraints of secrecy
and the reliability outage probabilities and transmit power budget
constraints. To solve this problem, we derived the closed-form optimal
RS rate expression of each user. Then, the $\epsilon$-constraint
transmit power of each user was calculated by the classic SPCA technique
under a given decoding order and then the optimal decoding order was
found by an exhaustive search method. Numerical results demonstrated
that the proposed algorithm significantly improves the effective network
secrecy throughput compared to the PD-NOMA and TDMA benchmarks, as
well as to the RSMA transmission ignoring CSIT uncertainty.

\appendices{}

\section{DERIVATION OF (\ref{eq:11})}

Based on (\ref{eq:8}), $P_{m,k,n}^{CO}$ can be expressed as $A1$,
\begin{figure*}[tbh]
\noindent \begin{centering}
{\small{}}%
\begin{tabular}{>{\centering}p{16cm}c}
\raggedright{}{\small{}$\mathbb{P}\left\{ 2^{r_{m,k,n}}-1>\frac{p_{m,k,n}\mathrm{PL}\left(d_{m,k}\right)\left|\mathbf{w}_{m}^{H}\mathbf{f}_{m,k}\right|^{2}}{\sum_{\left(k',n'\right)\in\Phi_{m,k,n}}p_{m,k',n'}\mathrm{PL}\left(d_{m,k'}\right)\,\left|\mathbf{w}_{m}^{H}\mathbf{f}_{m,k'}\right|^{2}+\stackrel[i=1,i\neq m]{M}{\sum}\,\stackrel[k''=1]{K_{i}}{\sum}P_{i,k''}\mathrm{PL}\left(d_{i,k''}\right)\left\Vert \sin\left(\phi_{i,k''}\right)\mathbf{f}_{i,k''}\right\Vert ^{2}\left|\mathbf{w}_{m}^{H}\mathbf{e}_{i,k''}\right|^{2}+\sigma_{m}^{2}}\right\} $} & {\small{}~~~~~~~}\tabularnewline
\end{tabular}{\small\par}
\par\end{centering}
\noindent \centering{}{\small{}}%
\begin{tabular}{>{\centering}p{16cm}c}
\raggedright{}{\small{}=$\mathbb{P}\left\{ 2^{r_{m,k,n}}-1>\frac{p_{m,k,n}\mathrm{PL}\left(d_{m,k}\right)\mathcal{X}_{k}}{\sum_{\left(k',n'\right)\in\Phi_{m,k,n}}p_{m,k',n'}\mathrm{PL}\left(d_{m,k'}\right)\mathcal{X}_{k'}+\stackrel[i=1,i\neq m]{M}{\sum}\,\stackrel[k''=1]{K_{i}}{\sum}P_{i,k''}\mathrm{PL}\left(d_{i,k''}\right)\mathcal{Y}_{i,k''}+\sigma_{m}^{2}}\right\} ,$} & {\small{}$\left(A1\right)$}\tabularnewline
\hline 
\end{tabular}{\small\par}
\end{figure*}
where we have $\mathcal{X}_{k}\triangleq\left|\mathbf{w}_{m}^{H}\mathbf{f}_{m,k}\right|^{2}\geq0$,
$\mathcal{Y}_{i,k"}\triangleq\left\Vert \mathbf{f}_{i,k''}\right\Vert ^{2}\sin^{2}\left(\phi_{i,k''}\right)\left|\mathbf{w}_{m}^{H}\mathbf{e}_{i,k''}\right|^{2}\geq0$.
Otherwise, $P_{m,k}^{CO}$ is always one. Furthermore, based on the
independence of the interference terms \cite[eq 24]{=00005B6=00005D},
the variables $\left\{ \mathcal{X}_{k}\right\} $ and $\left\{ \mathcal{Y}_{i,k''}\right\} $
$\forall i,k,k''$ are indeed independent.

To obtain a closed-form expression of $P_{m,k}^{CO}$ , we first provide
the probability density function (pdf) of $\mathcal{X}_{k}$ . Recall
that $\mathbf{n}_{m,k}\sim\mathcal{CN}(0,\,\mathbf{I}_{N_{t}})$ and
$\widetilde{\mathbf{f}}_{m,k}\triangleq\frac{\mathbf{f}_{m,k}}{\left\Vert \mathbf{f}_{m,k}\right\Vert }$,
$\mathcal{X}_{k}$ can be rewritten as $\mathcal{X}_{k}=\left\Vert \mathbf{f}_{m,k}\right\Vert ^{2}\left|\mathbf{w}_{m}^{H}\widetilde{\mathbf{f}}_{m,k}\right|^{2}$.
Since the normalized beamformer weights $\mathbf{w}_{m}$ are determined
by $\left\{ \mathbf{v}_{i}\right\} _{i=1,\,i\neq m}^{M}$ according
to $\mathbf{w}_{m}^{H}\mathbf{v}_{l}=0,\,\forall l\neq m,$ and $\left\{ \mathbf{v}_{i}\right\} _{i=1,\,i\neq m}^{M}$
are independent of $\mathbf{f}_{m,k}$, the vectors $\mathbf{f}_{m,k}$
and $\mathbf{w}_{m}$ are also independent. As a result, $\mathbf{f}_{m,k}$
and $\mathbf{w}_{m}$ are independent unit-norm vectors in the $N_{t}$-dimensional
space. Based on \cite[Lemma 1]{=00005B25=00005D}, the square inner
product between two independent unit-norm random vectors is Beta distributed
with shape parameters of $\left(1,N_{t}-1\right)$, i.e., we have
$X_{1}\triangleq\left|\mathbf{w}_{m}^{H}\widetilde{\mathbf{f}}_{m,k}\right|^{2}$
$\sim\textrm{Beta}\left(1,N_{t}-1\right)$ and its pdf is $f_{X_{1}}\left(x_{1}\right)=\frac{\left(1-x_{1}\right)^{N_{t}-2}}{\textrm{Be}\left(1,N_{t}-1\right)}$,
$x_{1}\in\left[0,1\right]$ \cite[ eq 8.380]{=00005B26=00005D}. On
the other hand, since we have $\mathbf{f}_{m,k}\sim\mathcal{CN}(0,\,\mathbf{I}_{N_{t}})$,
$X_{2}\triangleq\left\Vert \mathbf{f}_{m,k}\right\Vert ^{2}$ is distributed
as a chi-squared r.v. with $2N_{t}$ degrees of freedom as $\chi_{2N_{t}}^{2}$,
and its pdf is $f_{X_{2}}\left(x_{2}\right)=\frac{x_{2}^{N_{t}-1}e^{-\frac{x_{2}}{2}}}{2^{N_{t}}\varGamma\left(N_{t}\right)}$,
$x_{2}\geq0$, where $\varGamma\left(x\right)$ is the Gamma function
{[}50, eq. 8.310{]}. Since $\mathcal{X}_{k}=X_{1}X_{2}$, and $X_{1}$
and $X_{2}$ are independent, the pdf of $\mathcal{X}_{k}$ is given
by:
\begin{align*}
f_{\mathcal{X}_{k}}\left(x\right) & =\varint_{x}\frac{1}{\left|x_{2}\right|}f_{X_{2}}\left(x_{2}\right)f_{X_{1}}\left(\frac{x}{x_{2}}\right)dx_{2}\\
\, & =\frac{\varint_{x}^{+\infty}\left(x_{2}-x\right)^{N_{t}-2}e^{-\frac{x_{2}}{2}}dx_{2}}{\textrm{Be}\left(1,N_{t}-1\right)2^{N_{t}}\varGamma\left(N_{t}\right)}\\
\, & =\frac{1}{2}e^{-\frac{x}{2}},x\geq0=\textrm{Exp}(\frac{1}{2}).\,\,\,\,\,\,\,\,\,\,\,\,\,\,\,\,\,\,\,\,(A2)
\end{align*}

Thus, $\mathcal{X}_{k}\sim\textrm{Exp}(\frac{1}{2})$ is exponentially
distributed with rate $\lambda_{k}=\frac{1}{2}$. Now, we derive the
pdf of $\mathcal{Y}_{i,k''}$. The cumulative distribution function
of $\varOmega_{i,k"}\triangleq\sin^{2}\left(\phi_{i,k''}\right)$
is given by \cite{=00005B27=00005D}:{\small{}
\[
F_{\varOmega_{i,k"}}\left(\varpi\right)=\begin{cases}
\begin{array}{c}
1\end{array}, & \textrm{if}\,\,\,0\leq\varpi\leq2^{-\frac{B}{N_{t}-1}},\,\left(A3\right)\\
2^{B}\left(\varpi\right)^{N_{t}-1}, & \textrm{if}\,\,\,\varpi\geq2^{-\frac{B}{N_{t}-1}}.
\end{cases}
\]
}{\small\par}

Hence, we have $\left\Vert \mathbf{f}_{i,k"}\right\Vert ^{2}\sin^{2}\left(\phi_{i,k"}\right)\sim\textrm{Gamma}\left(N_{t}-1,2^{-\frac{B}{N_{t}-1}}\right)$,
which is gamma distributed with a shape parameter of $N_{t}-1$ and
scale parameter of $2^{-\frac{B}{N_{t}-1}}$ \cite[Lemma 1]{=00005B27=00005D}.
On the other hand, $\mathbf{e}_{i,k"}$ is a unit vector that has
the same distribution as $\mathbf{f}_{i,k"}$. Moreover, the unit
vector $\mathbf{w}_{m}$ is isotropic within the $\left(N_{t}-1\right)$-dimensional
hyperplane and independent of $\mathbf{e}_{i,k"}$. Based on \cite[Lemma 2]{=00005B23=00005D},
we have $\left|\mathbf{w}_{m}^{H}\mathbf{e}_{i,k"}\right|^{2}\sim\textrm{Beta}\left(1,N_{t}\text{\textminus}2\right)$.
Therefore the product $\mathcal{Y}_{i,k"}=\textrm{Beta}\left(1,N_{t}\text{\textminus}2\right)\times\textrm{Gamma}\left(N_{t}-1,2^{-\frac{B}{N_{t}-1}}\right)$
is exponentially distributed as \cite[Lemma 1]{=00005B28=00005D},
$\mathcal{Y}_{i,k"}=\left\Vert \mathbf{f}_{i,k"}\right\Vert ^{2}\sin^{2}\left(\phi_{i,k"}\right)\left|\mathbf{w}_{m}^{H}\mathbf{e}_{i,k"}\right|^{2}\sim\textrm{Exp}\left(2^{\frac{B}{N_{t}-1}}\right)$.
If we define the new variables{\small{} $\mathcal{\bar{X}}_{m,k',n'}\sim\textrm{Exp}\left(\lambda_{m,k',n'}=\frac{1}{2p_{m,k',n'}\mathrm{PL}\left(d_{m,k'}\right)}\right)$},
{\small{}$\mathcal{\bar{\mathcal{Y}}}_{i,k''}\sim\textrm{Exp}\left(\lambda_{i,k"}=\frac{2^{\frac{B}{N_{t}-1}}}{P_{i,k"}\mathrm{PL}\left(d_{i,k''}\right)}\right)$
,} then the $P_{m,k,n}^{CO}$ can be expressed as $\left(A4\right)$,
\begin{figure*}[tbh]
{\small{}
\begin{align*}
P_{m,k,n}^{CO} & =1-\mathbb{P}\left\{ \beta_{m,n,k}\leq\frac{\mathcal{\bar{X}}_{m,k,n}}{\sum_{\left(k',n'\right)\in\Phi_{m,k,n}}\mathcal{\bar{X}}_{m,k',n'}+\stackrel[i=1,i\neq m]{M}{\sum}\,\stackrel[k"=1]{K_{i}}{\sum}\mathcal{\bar{\mathcal{Y}}}_{i,k"}+\sigma_{m}^{2}}\right\} \\
 & =1-\mathbb{P}\left\{ \mathcal{\bar{X}}_{m,k,n}\geq\beta_{m,n,k}\left(\sum_{\left(k',n'\right)\in\Phi_{m,k,n}}\mathcal{\bar{X}}_{m,k',n'}+\stackrel[i=1,i\neq m]{M}{\sum}\,\stackrel[k''=1]{K_{i}}{\sum}\mathcal{\bar{\mathcal{Y}}}_{i,k''}+\sigma_{m}^{2}\right)\right\} \\
 & \overset{(a)}{=}1-\mathbb{E}\left\{ \exp\left(-\frac{\beta_{m,n,k}\left(\sum_{\left(k',n'\right)\in\Phi_{m,k,n}}\mathcal{\bar{X}}_{m,k',n'}+\stackrel[i=1,i\neq m]{M}{\sum}\,\stackrel[k''=1]{K_{i}}{\sum}\mathcal{\bar{\mathcal{Y}}}_{i,k''}+\sigma_{m}^{2}\right)}{2}\right)\right\} \\
 & \overset{(b)}{=}1-\mathbb{E}\left\{ \exp\left(-\frac{\beta_{m,n,k}\sigma_{m}^{2}}{2}\right)\stackrel[\left(k',n'\right)\in\Phi_{m,k,n}]{}{\prod}\mathcal{L}_{\mathcal{\bar{X}}_{m,k',n'}}\left(\frac{\beta_{m,n,k}}{2}\right)\stackrel[i=1,i\neq m]{M}{\prod}\stackrel[k''=1]{K_{i}}{\prod}\mathcal{L}_{\mathcal{\bar{\mathcal{Y}}}_{i,k''}}\left(\frac{\beta_{m,n,k}}{2}\right)\right\} \\
 & =1-\exp\left(-\frac{\beta_{m,n,k}\sigma_{m}^{2}}{2}\right)\stackrel[\left(k',n'\right)\in\Phi_{m,k,n}]{}{\prod}\stackrel[i=1,i\neq m]{M}{\prod}\stackrel[k"=1]{K_{i}}{\prod}\left(\frac{\lambda_{m,k',n'}}{\lambda_{m,k',n'}+\frac{\beta_{m,n,k}}{2}}\right)\left(\frac{\lambda_{i,k''}}{\lambda_{i,k''}+\frac{\beta_{m,n,k}}{2}}\right),\,\,\,\,\,\,\,\,\,\,\,\,\,\left(A4\right)
\end{align*}
\_\_\_\_\_\_\_\_\_\_\_\_\_\_\_\_\_\_\_\_\_\_\_\_\_\_\_\_\_\_\_\_\_\_\_\_\_\_\_\_\_\_\_\_\_\_\_\_\_\_\_\_\_\_\_\_\_\_\_\_\_\_\_\_\_\_\_\_\_\_\_\_\_\_\_\_\_\_\_\_\_\_\_\_\_\_\_\_\_\_\_\_\_\_\_\_\_\_\_\_\_\_}
\end{figure*}
where {\small{}$\mathcal{L}_{X}\left(s\right)\triangleq\mathbb{E}\left\{ e^{-sx}\right\} $
}is the Laplace transform. Step $(a)$ is reached by conditioning
on the aggregate interference{\small{} $\sum_{\left(k',n'\right)\in\Phi_{m,k,n}}$
$\mathcal{\bar{X}}_{m,k',n'}$ $+$ $\stackrel[i=1,i\neq m]{M}{\sum}\,\stackrel[k''=1]{K_{i}}{\sum}\mathcal{\bar{\mathcal{Y}}}_{i,k''}$}
and $(b)$ by the independence of the interference terms.

\section{DERIVATION OF (\ref{eq:12})}

Based on (\ref{eq:9}), $P_{j,k,n}^{SO}$ can be expressed as $\left(B1\right)$,
\begin{figure*}[tbh]
{\small{}
\begin{align*}
P_{j,k,n}^{SO} & =\mathbb{P}\left\{ 2^{D_{j,k,n}}-1<\frac{p_{m,k,n}\mathrm{PL}\left(d_{m,k,j}\right)\left|g_{m,k,j}\right|^{2}}{\sum_{\left(k',n'\right)\neq\left(k,n\right)}p_{m,k',n'}\mathrm{PL}\left(d_{m,k',j}\right)\left|g_{m,k',j}\right|^{2}+\sigma_{e}^{2}}\right\} \,\,\,\,\,\,\,\,\,\,\,\,\,\,\,\,\,\,\,\,\,\,\,\,\,\,\,\,\,\,\,\,\,\,\,\,\,\,\,\,\,\,\,\,\,\,\,\,\,\,\,\,\,\,\,\,\,\,\,\,\,\,\,\,\,\,\,\,\,\,\,\,\,\,\,\,\left(B1\right)\\
 & =\mathbb{P}\left\{ 2^{D_{j,k,n}}-1<\frac{\mathcal{W}_{j,k,n}}{\sum_{\left(k',n'\right)\neq\left(k,n\right)}\mathcal{U}_{j,k',n'}+\sigma_{e}^{2}}\right\} ,\,\,
\end{align*}
\_\_\_\_\_\_\_\_\_\_\_\_\_\_\_\_\_\_\_\_\_\_\_\_\_\_\_\_\_\_\_\_\_\_\_\_\_\_\_\_\_\_\_\_\_\_\_\_\_\_\_\_\_\_\_\_\_\_\_\_\_\_\_\_\_\_\_\_\_\_\_\_\_\_\_\_\_\_\_\_\_\_\_\_\_\_\_\_\_\_\_\_\_\_\_\_\_\_\_\_\_\_}
\end{figure*}
where $\mathcal{W}_{j,k,n}\triangleq p_{m,k,n}\mathrm{PL}\left(d_{m,k,j}\right)\lambda_{j}^{max}\left\{ \tilde{\mathcal{\mathbf{Q}}}_{m,j}\right\} \left|g_{m,k,j}\right|^{2}$
and $\mathcal{U}_{j,k',n'}\triangleq p_{m,k',n'}\mathrm{PL}\left(d_{m,k',j}\right)\lambda_{j}^{max}\left\{ \tilde{\mathcal{\mathbf{Q}}}_{m,j}\right\} \left|g_{m,k',j}\right|^{2}$
are independent exponential r.v obeying the distribution of $\mathcal{W}_{j,k,n}\sim\textrm{Exp}\left(\eta_{j,k,n}=\frac{1}{p_{m,k,n}\mathrm{PL}\left(d_{m,k,j}\right)\lambda_{j}^{max}\left\{ \tilde{\mathcal{\mathbf{Q}}}_{m,j}\right\} }\right)$
and $\mathcal{U}_{k',n}\sim\textrm{Exp}\left(\zeta_{j,k',n'}=\frac{1}{p_{m,k',n'}\mathrm{PL}\left(d_{m,k',j}\right)\lambda_{j}^{max}\left\{ \tilde{\mathcal{\mathbf{Q}}}_{m,j}\right\} }\right)$.
Then the $P_{j,k,n}^{SO}$ can be expressed as $\left(B2\right)$,
\begin{figure*}[tbh]
{\footnotesize{}
\begin{align*}
P_{j,k,n}^{SO} & =\mathbb{P}\left\{ \mathcal{W}_{j,k,n}>\kappa_{j,k,n}\left(\sum_{\left(k',n'\right)\neq\left(k,n\right)}\mathcal{U}_{j,k',n'}+\sigma_{e}^{2}\right)\right\} \overset{(a)}{=}\mathbb{P}\left\{ \exp\left(-\eta_{j,k,n}\kappa_{j,k,n}\left(\sum_{\left(k',n'\right)\neq\left(k,n\right)}\mathcal{U}_{j,k',n'}+\sigma_{e}^{2}\right)\right)\right\} ,\,\,\,\left(B2\right)\\
 & \overset{(b)}{=}\mathbb{E}\left\{ \exp\left(-\eta_{j,k,n}\kappa_{j,k,n}\sigma_{e}^{2}\right)\stackrel[\left(k',n'\right)\neq\left(k,n\right)]{}{\prod}\mathcal{L}_{\mathcal{U}_{k',n}}\left(\eta_{j,k,n}\kappa_{j,k,n}\right)\right\} =\exp\left(-\eta_{j,k,n}\kappa_{j,k,n}\sigma_{e}^{2}\right)\stackrel[\left(k',n'\right)\neq\left(k,n\right)]{}{\prod}\left(1+\eta_{j,k,n}\kappa_{j,k,n}\zeta_{j,k',n'}^{-1}\right)^{-1},
\end{align*}
}{\small{}\_\_\_\_\_\_\_\_\_\_\_\_\_\_\_\_\_\_\_\_\_\_\_\_\_\_\_\_\_\_\_\_\_\_\_\_\_\_\_\_\_\_\_\_\_\_\_\_\_\_\_\_\_\_\_\_\_\_\_\_\_\_\_\_\_\_\_\_\_\_\_\_\_\_\_\_\_\_\_\_\_\_\_\_\_\_\_\_\_\_\_\_\_\_\_\_\_\_\_\_\_\_}
\end{figure*}
where $\kappa_{j,k,n}\triangleq2^{D_{j,k,n}}-1$.

\section{DERIVATION OF (\ref{eq:14})}

Upon using the inequality of $\frac{1}{1+x}\leq\frac{1}{x}$ we have
$\left(C1\right)$ and $\left(C2\right)$. 
\begin{figure*}[tbh]
\begin{centering}
{\small{}}%
\begin{tabular}{>{\centering}p{16cm}c}
{\small{}$\exp\left(-\eta_{j,k,n}\kappa_{j,k,n}\sigma_{e}^{2}\right)\stackrel[\left(k',n'\right)\neq\left(k,n\right)]{}{\prod}\left(1+\eta_{j,k,n}\kappa_{j,k,n}\zeta_{j,k',n'}^{-1}\right)^{-1}\overset{}{\leq}\Psi\left(\kappa_{j,k,n}\right)\leq{\epsilon_{sop}},$} & {\small{}$\left(C1\right)$}\tabularnewline
\end{tabular}{\small\par}
\par\end{centering}
\begin{centering}
{\small{}}%
\begin{tabular}{>{\centering}p{16cm}c}
{\small{}$\Psi\left(\kappa_{j,k,n}\right)\triangleq\exp\left(-\eta_{j,k,n}\kappa_{j,k,n}\sigma_{e}^{2}\right)\kappa_{j,k,n}^{-\left(2K_{m}-1\right)}\frac{\stackrel[\left(k',n'\right)\neq\left(k,n\right)]{}{\prod}\zeta_{j,k',n'}}{\eta_{j,k,n}}.$} & {\small{}$\left(C2\right)$}\tabularnewline
\end{tabular}{\small\par}
\par\end{centering}
\begin{centering}
{\small{}}%
\begin{tabular}{>{\centering}p{16cm}c}
{\small{}$\Psi^{'}\left(\kappa_{j,k,n}\right)=-\kappa_{j,k,n}^{-\left(2K_{m}-1\right)}\left(\eta_{j,k,n}\sigma_{e}^{2}+\frac{\left(2K_{m}-1\right)}{\kappa_{j,k,n}}\right)<0,\,\,\kappa_{j,k,n}>0,$} & {\small{}$\left(C3\right)$}\tabularnewline
\end{tabular}{\small\par}
\par\end{centering}
\begin{centering}
{\small{}}%
\begin{tabular}{>{\centering}p{16cm}c}
{\small{}$\exp\left(-\eta_{j,k,n}\kappa_{j,k,n}\sigma_{e}^{2}\right)\kappa_{j,k,n}^{-\left(2K_{m}-1\right)}\frac{\stackrel[\left(k',n'\right)\neq\left(k,n\right)]{}{\prod}\zeta_{j,k',n'}}{\eta_{j,k,n}}={\epsilon_{sop}}.$} & {\small{}$\left(C4\right)$}\tabularnewline
\end{tabular}{\small\par}
\par\end{centering}
\begin{centering}
{\small{}}%
\begin{tabular}{>{\centering}p{16cm}c}
{\small{}$\frac{\eta_{j,k,n}\sigma_{e}^{2}}{\left(2K_{m}-1\right)}\kappa_{j,k,n}\exp\left(-\frac{\eta_{j,k,n}\sigma_{e}^{2}}{\left(2K_{m}-1\right)}\kappa_{j,k,n}\right)=\frac{\eta_{j,k,n}\sigma_{e}^{2}}{\left(2K_{m}-1\right)}\left(\frac{\stackrel[\left(k',n'\right)\neq\left(k,n\right)]{}{\prod}\zeta_{j,k',n'}}{\eta_{j,k,n}}{\epsilon_{sop}}^{-1}\right)^{\frac{1}{\left(2K_{m}-1\right)}}.$} & {\small{}$\left(C5\right)$}\tabularnewline
\end{tabular}{\small\par}
\par\end{centering}
\centering{}{\small{}}%
\begin{tabular}{>{\centering}p{16cm}c}
{\small{}$W_{0}\left(\frac{\eta_{j,k,n}\sigma_{e}^{2}}{\left(2K_{m}-1\right)}\left(\frac{\stackrel[\left(k',n'\right)\neq\left(k,n\right)]{}{\prod}\zeta_{j,k',n'}}{\eta_{j,k,n}}{\epsilon_{sop}}^{-1}\right)^{\frac{1}{\left(2K_{m}-1\right)}}\right)=\frac{\eta_{j,k,n}\sigma_{e}^{2}}{\left(2K_{m}-1\right)}\kappa_{j,k,n}.$} & {\small{}$\left(C6\right)$}\tabularnewline
\hline 
\end{tabular}{\small\par}
\end{figure*}

The first-order derivative of $\Psi\left(\kappa_{j,k,n}\right)$ is
given by $\left(C3\right)$, which is negative. By exploiting the
decreasing nature of $\Psi\left(\kappa_{j,k,n}\right)$ and its lower-bounded
nature in $\left(C2\right)$, the minimum value of $D_{j,k,n}$, is
obtained by solving the $\left(C4\right)$. By straightforward algebra,
($C4$) can be further reformulated as $\left(C5\right)$. Finally,
with the help of the principal branch of the Lambert $W$-function
\cite{=00005B18=00005D}, ($C5$) is rewritten as $\left(C6\right)$.
Re-arranging the terms in ($C6$) leads to (\ref{eq:14}).


\begin{thebibliography}{10}
\bibitem{=00005B1=00005D} M. Alzenad, A. El-Keyi, and H. Yanikomeroglu,
\textquotedblleft 3-D placement of an unmanned aerial vehicle base
station for maximum coverage of users with different QoS requirements,\textquotedblright{}
\textit{IEEE Wireless Communications Letters}, vol. 7, no. 1, pp.
38\textendash 41, 2017. 

\bibitem{=00005B1-1=00005D}Y. Sun, D. Xu, D. W. K. Ng, L. Dai and
R. Schober, \textquotedbl Optimal 3D-Trajectory Design and Resource
Allocation for Solar-Powered UAV Communication Systems,\textquotedbl{}
\textit{in IEEE Transactions on Communications}, vol. 67, no. 6, pp.
4281-4298, June 2019, doi: 10.1109/TCOMM.2019.2900630.

\bibitem{=00005B2=00005D} W. Jaafar, N. Shimaa, M. Sami, C.S Paschalis,
and H. Yanikomeroglu. \textquotedbl On the downlink performance of
RSMA-based UAV communications.\textquotedbl{} \textit{IEEE Transactions
on Vehicular Technology 69}, no. 12 (2020): 16258-16263. 

\bibitem{=00005B3=00005D}H. Zhang, J. Zhang and K. Long, \textquotedbl Energy
Efficiency Optimization for NOMA UAV Network with Imperfect CSI,\textquotedbl{}
in \textit{IEEE Journal on Selected Areas in Communications}, vol.
38, no. 12, pp. 2798-2809, Dec. 2020, doi: 10.1109/JSAC.2020.3005489. 

\bibitem{=00005B4=00005D}H. Bastami, M. Letafati, M. Moradikia, A.
Abdelhadi, H. Behroozi and L. Hanzo, \textquotedbl On the Physical
Layer Security of the Cooperative Rate-Splitting Aided Downlink in
UAV Networks,\textquotedbl{} \textit{in IEEE Transactions on Information
Forensics and Security}, doi: 10.1109/TIFS.2021.312298 

\bibitem{=00005B5=00005D} H. Bastami, M. Moradikia, M. Letafati,
A. Abdelhadi, and H. Behroozi. \textquotedbl Outage-Constrained Robust
and Secure Design for Downlink Rate-Splitting UAV Networks.\textquotedbl{}
\textit{In 2021 IEEE International Conference on Communications Workshops
(ICC Workshops)}, pp. 1-7. IEEE, 2021. 

\bibitem{=00005B6=00005D}M. Kountouris, and J. G. Andrews. \textquotedbl Downlink
SDMA with limited feedback in interference-limited wireless networks.\textquotedbl{}
\textit{IEEE Transactions on Wireless Communications 11}, no. 8 (2012):
2730-2741. 

\bibitem{=00005B7=00005D} Z. Li, M. Xia, M. Wen, and Y. Wu. \textquotedbl Massive
access in secure NOMA under imperfect CSI: security guaranteed sum-rate
maximization with first-order algorithm.\textquotedbl{} \textit{IEEE
Journal on Selected Areas in Communications 39}, no. 4 (2020): 998-1014. 

\bibitem{=00005B8=00005D}Y. Mao, B. Clerckx, and V. O. Li, \textquotedblleft Rate-splitting
multiple access for downlink communication systems: bridging, generalizing,
and outperforming SDMA and NOMA,\textquotedblright{} \textit{EURASIP
journal on wireless communications and networking}, vol. 2018, no.
1, p. 133, 2018. 

\bibitem{=00005B9=00005D}H. Fu, S. Feng, W. Tang and D. W. K. Ng,
\textquotedbl Robust Secure Beamforming Design for Two-User Downlink
MISO Rate-Splitting Systems,\textquotedbl{} \textit{in IEEE Transactions
on Wireless Communications}, vol. 19, no. 12, pp. 8351- 8365, Dec.
2020. 

\bibitem{=00005B10=00005D}J. Zeng, T. Lv, W. Ni, R. P. Liu, N. C.
Beaulieu and Y. J. Guo, \textquotedbl Ensuring Max\textendash Min
Fairness of UL SIMO-NOMA: A Rate Splitting Approach,\textquotedbl{}
\textit{in IEEE Transactions on Vehicular Technology,} vol. 68, no.
11, pp. 11080-11093, Nov. 2019, doi: 10.1109/TVT.2019.2943511.

\bibitem{=00005B11=00005D}O. Abbasi, and H. Yanikomeroglu, ``Rate-Splitting
and NOMA-Enabled Uplink User Cooperation.'' \textit{In 2021 IEEE
Wireless Communications and Networking Conference Workshops (WCNCW)
}(pp. 1-6). IEEE, 2021.

\bibitem{=00005B12=00005D}Z. Yang, M. Chen, W. Saad, W. Xu, and M.
Shikh-Bahaei. \textquotedbl Sum-rate maximization of uplink rate
splitting multiple access (RSMA) communication.\textquotedbl{} \textit{IEEE
Transactions on Mobile Computing (2020)}. 

\bibitem{=00005B13=00005D} S. -. Lin, T. -. Chang, Y. -. Hong and
C. -. Chi, \textquotedbl On the Impact of Quantized Channel Direction
Feedback in Multiple-Antenna Wiretap Channels,\textquotedbl{} \textit{2010
IEEE International Conference on Communications,} 2010, pp. 1-5, doi:
10.1109/ICC.2010.5502608.

\bibitem{=00005B14=00005D}M. Moradikia, H. Bastami, A. Kuhestani,
H. Behroozi, and L. Hanzo, \textquotedblleft Cooperative secure transmission
relying on optimal power allocation in the presence of untrusted relays,
a passive eavesdropper and hardware impairments,\textquotedblright{}
\textit{IEEE Access}, vol. 7, pp. 116 942\textendash 116 964, 2019. 

\bibitem{=00005B15=00005D}M. Moradikia, S. Mashdour, and A. Jamshidi.
\textquotedbl Joint optimal power allocation, cooperative beamforming,
and jammer selection design to secure untrusted relaying network.\textquotedbl{}
\textit{Transactions on Emerging Telecommunications Technologies 29},
no. 3 (2018): e3276. 

\bibitem{=00005B16=00005D}Y. Sun, D. W. K. Ng, J. Zhu and R. Schober,
\textquotedbl Robust and Secure Resource Allocation for Full-Duplex
MISO Multicarrier NOMA Systems,\textquotedbl{} \textit{in IEEE Transactions
on Communications}, vol. 66, no. 9, pp. 4119-4137, Sept. 2018, doi:
10.1109/TCOMM.2018.2830325.

\bibitem{=00005B17=00005D}Y. Li, M. Jiang, Q. Zhang, Q. Li and J.
Qin, \textquotedbl Secure Beamforming in Downlink MISO Nonorthogonal
Multiple Access Systems,\textquotedbl{} \textit{in IEEE Transactions
on Vehicular Technology,} vol. 66, no. 8, pp. 7563-7567, Aug. 2017,
doi: 10.1109/TVT.2017.2658563.

\bibitem{=00005B18=00005D}R. M. Corless, G. H. Gonnet, D. E. G. Hare,
D. J. Jeffrey, and D. E. Knuth, \textquotedblleft On the Lambert W
function,\textquotedblright{} \textit{Advances in Computational Mathematics},
vol. 5, no. 1, pp. 329\textendash 359, Dec. 1996.

\bibitem{=00005B19=00005D}A. Omri and M. O. Hasna, \textquotedbl Physical
Layer Security Analysis of UAV Based Communication Networks,\textquotedbl{}
\textit{2018 IEEE 88th Vehicular Technology Conference (VTC-Fall)},
2018, pp. 1-6, doi: 10.1109/VTCFall.2018.8690950.

\bibitem{19_} B. Mehlig and J. T. Chalker, ``Statistical properties
of eigenvectors in non-Hermitian Gaussian random matrix ensembles'',
\textit{Journal of Mathematical Physics}, 2000, https://doi.org/10.1063\%2F1.533302.

\bibitem{=00005B20=00005D}B. Clerckx and C. Oestges, \textit{MIMO
Wireless Networks: Channels, Techniques and Standards for Multi-Antenna,
Multi-User and Multi-Cell Systems}, 2nd ed. Academic Press, 2013.

\bibitem{=00005B21=00005D}D. J. Love, R. W. Heath and T. Strohmer,
\textquotedbl Grassmannian beamforming for multiple-input multiple-output
wireless systems,\textquotedbl{} \textit{in IEEE Transactions on Information
Theory}, vol. 49, no. 10, pp. 2735-2747, Oct. 2003, doi: 10.1109/TIT.2003.817466.

\bibitem{=00005B22=00005D}K. K. Mukkavilli, A. Sabharwal, E. Erkip
and B. Aazhang, \textquotedbl On beamforming with finite rate feedback
in multiple-antenna systems,\textquotedbl{} \textit{in IEEE Transactions
on Information Theory}, vol. 49, no. 10, pp. 2562-2579, Oct. 2003,
doi: 10.1109/TIT.2003.817433.

\bibitem{=00005B23=00005D}N. Jindal, \textquotedblleft MIMO broadcast
channels with finite-rate feedback,\textquotedblright{}\textit{ IEEE
Trans. Inf. Theory}, vol. 52, no. 11, pp. 5045\textendash 5060, Nov.
2006.

\bibitem{=00005B24=00005D}B. Rimoldi and R. Urbanke, \textquotedbl A
rate-splitting approach to the Gaussian multiple-access channel,\textquotedbl{}
\textit{in IEEE Transactions on Information Theory,} vol. 42, no.
2, pp. 364-375, March 1996, doi: 10.1109/18.485709.

\bibitem{=00005B25=00005D}J. C. Roh and B. D. Rao, \textquotedbl Transmit
beamforming in multiple-antenna systems with finite rate feedback:
a VQ-based approach,\textquotedbl\textit{ in IEEE Transactions on
Information Theory,} vol. 52, no. 3, pp. 1101-1112, March 2006, doi:
10.1109/TIT.2005.864426.

\bibitem{25_1} W. Wang, K. C. Teh, and K. H. Li, \textquotedblleft Secrecy
throughput maximiza- tion for MISO multi-eavesdropper wiretap channels,\textquotedblright{}
\textit{IEEE Trans. Inf. Forensics Security}, vol. 12, no. 3, pp.
505\textendash 515, Mar. 2017.

\bibitem{=00005B26=00005D}A. Jeffrey and D. Zwillinger, \textit{Table
of Integrals, Series, and Products (6th ed.)}. San Diego, USA: Academic
Press, 2000.

\bibitem{=00005B27=00005D}T. Yoo, N. Jindal and A. Goldsmith, \textquotedbl Multi-Antenna
Downlink Channels with Limited Feedback and User Selection,\textquotedbl{}
\textit{in IEEE Journal on Selected Areas in Communications,} vol.
25, no. 7, pp. 1478-1491, September 2007, doi: 10.1109/JSAC.2007.070920.

\bibitem{=00005B28=00005D}J. Zhang, R. W. Heath, M. Kountouris, and
J. G. Andrews, \textquotedblleft Mode switching for the multi-antenna
broadcast channel based on delay and channel quantization,\textquotedblright{}
\textit{EURASIP J. Adv. Sig. Proc., }vol. 2009, no. 1, p. 802548,
Jun. 2009.

\bibitem{=00005B30=00005D}S. Boyd and L. Vandenberghe, Convex optimization.
\textit{Cambridge university press}, 2004.

\bibitem{=00005B31=00005D}H. Lei \textit{et al}., \textquotedbl On
Secure Mixed RF-FSO Systems With TAS and Imperfect CSI,\textquotedbl\textit{
in IEEE Transactions on Communications}, vol. 68, no. 7, pp. 4461-4475,
July 2020, doi: 10.1109/TCOMM.2020.2985028.

\bibitem{34} H. Bastami, M. Moradikia, H. Behroozi, R. C. de Lamare,
A. Abdelhadi, Z. Ding, ``Secrecy rate maximization for hardware impaired
untrusted relaying network with deep learning'', \textit{Physical
Communication}, Volume 49, 2021, 101476, ISSN 1874-4907, https://doi.org/10.1016/j.phycom.2021.101476. 

\bibitem{35} S. Boyd and L. Vandenberghe, Convex Optimization. USA:
\textit{Cambridge University Press}, 2004.

\bibitem{36} D. Bertsekas, Nonlinear Programming. Athena Scientific,
1999.
\end{thebibliography}
\end{document}